\documentclass[journal]{IEEEtran}
\usepackage{booktabs}
\usepackage{xcolor}
\usepackage{colortbl}
\usepackage{amssymb}
\usepackage{algorithmic}
\usepackage{amsmath} 
\usepackage{amsfonts}
\newtheorem{theorem}{Theorem}

\newtheorem{proposition}[theorem]{Proposition}

\newtheorem{definition}[theorem]{Definition}
\newtheorem{remark}[theorem]{Remark}
\newtheorem{problem}[theorem]{Problem}
\newtheorem{algorithm}[theorem]{Algorithm}

\usepackage{graphicx}
\usepackage{array}

\ifCLASSOPTIONcompsoc
  \usepackage[caption=false,font=normalsize,labelfont=sf,textfont=sf]{subfig}
\else
  \usepackage[caption=false,font=footnotesize]{subfig}
\fi
\usepackage{float}
\usepackage{fixltx2e}
\usepackage{stfloats}
\usepackage{color}
\usepackage{xcolor}
\usepackage{textcomp}
\usepackage[normalem]{ulem}

\IEEEoverridecommandlockouts

\begin{document}
\title{Hybrid Controller for Wind Turbine Generators to Ensure Adequate Frequency Response in Power Networks}

\author{Yichen~Zhang,~\IEEEmembership{Student Member,~IEEE,}
        Kevin~Tomsovic,~\IEEEmembership{Fellow,~IEEE,}
        ~Seddik~M.~Djouadi,~\IEEEmembership{Member,~IEEE}
        and~Hector~Pulgar-Painemal,~\IEEEmembership{Member,~IEEE}% <-this % stops a space
\thanks{This work was supported in part by the National Science Foundation under grant No CNS-1239366, in part by the National Science Foundation under grant No ECCS-1509114, and in part by the Engineering Research Center Program of the National Science Foundation and the Department of Energy under NSF Award Number EEC-1041877.}% <-this % stops a space
\thanks{The authors are with the Min H. Kao Department of Electrical Engineering and Computer Science, The University of Tennessee, Knoxville, TN 37996 USA (e-mail: yzhan124@utk.edu).}}% <-this % stops a space

\markboth{P\MakeLowercase{ublished on} IEEE JOURNAL ON EMERGING AND SELECTED TOPICS IN CIRCUITS AND SYSTEMS \MakeLowercase{in} S\MakeLowercase{eptember} (DOI: 10.1109/JETCAS.2017.2675879)}
{Shell \MakeLowercase{\textit{et al.}}: Bare Demo of IEEEtran.cls for IEEE Journals}
\maketitle
\begin{abstract}
Converter-interfaced power sources (CIPS) are hybrid control systems as they may switch between multiple operating modes. Due to increasing penetration, the hybrid behavior of CIPS, such as, wind turbine generators (WTG), may have significant impact on power system dynamics. In this paper, the frequency dynamics under inertia emulation and primary support from WTG is studied. A mode switching for WTG to ensure adequate frequency response is proposed. The switching instants are determined by our proposed concept of a region of safety (ROS), which is the initial set of safe trajectories. The barrier certificate methodology is employed to derive a new algorithm to obtain and enlarge the ROS for the given desired safe limits and the worst-case disturbance scenarios. Then critical switching instants and a safe recovery procedure are found. In addition, the emulated inertia and load-damping effect is derived in the time frame of inertia and primary frequency response, respectively. The theoretical results under critical cases are consistent with simulations and can be used as guidance for practical control design.
\end{abstract}
\begin{IEEEkeywords}
System frequency response, deadband, hybrid system, inertia emulation, primary frequency support, safety verification, sum of squares decomposition, semidefinite programming,  wind turbine generator.
\end{IEEEkeywords}
\IEEEpeerreviewmaketitle

\section{Introduction}
\IEEEPARstart{H}ybrid behaviors in complex power networks have not been carefully studied. However, with the increasing connection of converter-interfaced power sources (CIPS), such as, wind turbine generators (WTGs), into the power grid, complex switching behaviors have been introduced as CIPS can operate in many  different modes such as grid-feeding, grid-forming and grid-supporting \cite{mg_control}. The complex hybrid behaviors from integrated CIPS will have more significant impact on the traditional grid due to increasing penetration \cite{wind_report}. Analysis of the existing modes and corresponding switchings is important to understand system limits and guidelines for control design.\par
Wind power is a dominant source among all renewable sources. Variable speed wind turbine generators (WTG) are mechanically decoupled from the grid and do not automatically respond to frequency changes. With increasing penetration of wind power, the natural frequency support of traditional synchronous generators is decreasing. Configuring the WTG controls to participate in frequency control could benefit power system dynamics by reducing frequency excursions \cite{wu_2013_towards}. This participation can be realized by adding supplementary control loops to the normal maximum power point tracking (MPPT) mode of the WTG as shown in Fig. \ref{fig_WTG_control} \cite{morren2006wind}.\par
Wind farm should operate at MPPT mode during most times for efficient energy extraction, but provide enough active power to form a synthetic inertia during certain events to ensure system frequency stay within safe limits to avoid triggering protection \cite{load_shed}. Such \emph{performance guaranteed control} concept have proposed as a new objective for highly controllable converters \cite{impact_ETH} \cite{rate_MIT}. The physical component corresponds to this hybrid dynamics is a deadband. It is necessary for efficient operation by guaranteeing more power extraction from the wind and less mechanical stress on the gearbox; however, a large deadband may limit the opportunity for the WTG to provide sufficient inertia during a disturbance \cite{ge_report}. This is a crucial trade-off between economics and reliable operations.\par
The aforementioned issues lead to the following two questions: Under a certain disturbance, can the designed inertia emulation preserve the desired frequency limits? If so, what is the largest deadband that preserves these limits? These questions arise from actual power system operations faced by transmission system operators (TSO) such as, Hydro-Qu{\'e}bec (HQ) \cite{quebec}. However, as pointed out in \cite{rate_MIT}, available time (equivalent to deadband setting) for CIPS to maintain bounded frequency is usually unknown. Few methods have been proposed to answer the above questions beyond extensive simulations.\par
In this paper, we propose a systematic theoretical analysis to find precise answers to the above questions by considering hybrid dynamic models. Based on selected modal analysis (SMA) \cite{hector}, a computationally truncable reduced-order model is obtained and the above questions become tractable and solvable based on the barrier certificate approaches for hybrid system safety verification \cite{barrier}. In addition, the synthetic inertia and damping provided by CIPS is derived based on the reduced-order model.\par
\subsection{Related works}
\subsubsection{Hybrid system safety verification}
\begin{figure}[t]
	\centering
	\includegraphics[scale=0.52]{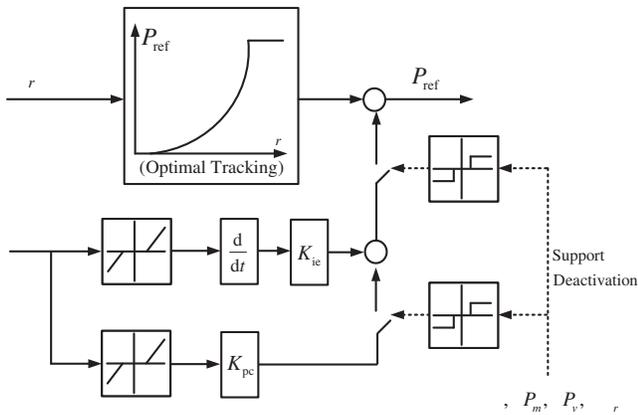}
	\caption{Active power control of wind turbine generator with inertia emulation and primary frequency control.}
	\label{fig_WTG_control}
\end{figure}

\begin{figure}[t]
	\centering
	\includegraphics[scale=0.54]{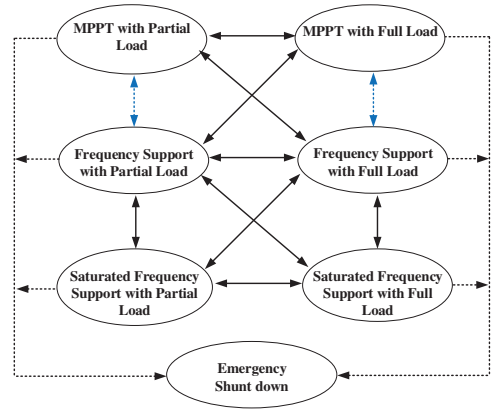}
	\caption{Mode transitions of a WTG.}
	\label{fig_WTG_transition}
\end{figure}

A hybrid system consists of continuous dynamic subsystems and discrete events that capture interactions between them. Safety verification of hybrid systems combines the automatic verification techniques for finite state concurrent systems (so-called model checking techniques) and computation of reachable sets for continuous dynamic systems \cite{hs_verify_survey}. The mode transitions of controller in Fig. \ref{fig_WTG_control} is shown in Fig. \ref{fig_WTG_transition}. Some switchings take place autonomously due to physical limits (solid lines), while others (dash lines) are designed for specific purposes, which are in the scope of this paper. Due to the relatively simple transition map, safety verification reduces down to reachable sets computation under different vector fields.\par
The different approaches in reachability analysis of continuous dynamic system can be categorized into Lagrangian and Eulerian methods \cite{safety_2_cato}. Lagrangian methods seek efficient over-approximation of the reachable set by propagating certain initial sets represented usually by polygons or ellipses under system vector field. Lagrangian methods are computationally feasible for high-dimensional systems, and have been successfully applied to large scale power systems \cite{ieee_14_ac} \cite{ieee_14_ps} \cite{renewable_variability}. These approximations lack accuracy when the shape of the reachable set is not a polygon or an ellipse. On the other hand, the goal of Eulerian methods (also known as level set method) is to calculate as closely as possible the true reachable set by computing a numerical solution to the Hamilton-Jacobi partial differential equation (HJ PDE), where the initial sets are implicitly represented by zero sublevel sets of an appropriate function. This is known as convergent approximation. Transient \cite{hs_power_survey} \cite{lsv_transient} and voltage \cite{lsv_voltage} stability can be precisely analyzed with the help of this method. To obtain numerical solutions to the HJ PDEs, one needs to discretize the state space, which leads to exponentially increasing computational complexity and limits its application to systems with no more than four continuous variables \cite{ieee_14_ac}.\par
If the system dynamics and safety specifications can be represented as polynomials, references \cite{sos} and \cite{barrier} propose a passivity-based approach that formulates safety verification as a sum of squares (SOS) optimization problem. As long as the SOS program is feasible, the safety property can be verified and a polynomial barrier certificate is obtained such that no trajectory of the system starting from the initial set can cross this barrier to reach an unsafe region. The formulation in \cite{barrier} leads to an arbitrary barrier certificate, which can not represent the maximum safety preserving capability. The safety supervisors for wind turbine emergency shutdown is designed in \cite{shutdown} where the conservatism is reduced by maximizing the volume of an elliptical reference shape inside the barrier certificate. Still, no trajectory evaluation is attempted and because of the shape limitation of the ellipsoid, it is difficult to attain satisfactory results. In addition, the formulations in \cite{barrier} and \cite{shutdown} require a specific initial set. In a hybrid system, the initial conditions after switching depend on the switching instant. This requires a framework that builds the condition without specific initial sets. Despite these issues, the framework provides useful flexibility between accuracy and computational complexity by choosing an appropriate polynomial order.
\subsubsection{Frequency control with participation of WTG considering deadband and safety limits}
Accurate modeling of deadband and other thresholds as hybrid systems can be done using piecewise linear approximation \cite{hiskens}. GE has proposed a detailed WTG model with inertia emulation and primary frequency controllers, where the deadband is included. A simplified structure is shown in Fig. \ref{fig_WTG_control} \cite{ge_report}. Recommended values are 0.15 Hz for inertia emulation and 0.24 Hz for primary frequency control, respectively, but the justification for these values is not clear. Based on this control configuration, a unified deadband of 0.1 Hz has been studied through simulation \cite{delft}. In \cite{quebec} Hydro-Qu{\'e}bec identified the need for quantifying inertia emulation with deadband under the safety requirement. In their approach, a frequency excursion limit of 1.5 Hz is set up to prevent load shedding, then a certain amount of inertia emulation and deadband are determined based on simulation studies. 
A second drop of grid frequency will occur once the primary frequency control of a partially loaded WTG is deactivated \cite{morren2006wind} \cite{ge_report} \cite{review_2_drop}. A soft recovery procedure, i.e., deactivating primary frequency control of different WTGs at different times, is proposed in \cite{end_at_diff} \cite{soft_recover}. The safety region has not been considered in this scenario.
\subsection{Our Contributions}
We propose the very concept \emph{region of safety} to clarify the hybrid system safety verification. A convergent approximation algorithm is developed to estimate the largest region of safety instead of achieving an arbitrary barrier certificate containing a specific initial set. From the practical engineering point of view, the established framework is applied to solve the hybrid mode synthesis problem. The switching criteria or equivalent deadbands for bounded response are built and explained. The framework and analysis in this paper provide a general guideline for the unclear safe switching problem.\par
\subsection{Outline}
The outline of the paper is as follows. In Section \ref{sec_framework}, the barrier certificate methodology and our proposed algorithm is introduced. In Section \ref{sec_model}, the concept of representing deadband as a hybrid systems is presented. Then selected modal analysis (SMA) based model reduction is employed. The controller gain is preserved under certain assumptions and their equivalent emulated inertia and load-damping coefficient is calculated. A case study is presented in Section \ref{sec_case_study} and followed by the conclusions in Section \ref{sec_con}.\par

\section{Principle of Safe Mode Switching Synthesis}
\label{sec_framework}
Safety denotes the property that all system trajectories stay within given bounded regions, thus, equipment damage or relay trigger can be avoided (Note this is similar, but not identical, to what is called security in power industry but for purposes of this paper we will assume satisfying safety conditions ensures secure operation). Consider the dynamics of a power system governed by a set of ordinary differential equations (ODEs) as
\begin{equation}
\label{eq_ode}
\dot{x}(t)=f(x(t),d(t))
\end{equation}
where $x(t) \in \mathbb{R}^{n}$ denotes the vector of state variables and $d(t)\in \mathbb{R}^{m}$ denotes certain disturbances, such as, generation loss or an abrupt load change. Such a disturbance may be assumed to be piecewise constant and bounded in the set $\mathcal{D}$. Let $\mathcal{X}\subseteq\mathbb{R}^{n}$ be the computational domain of interest, $\mathcal{X}_{I}\subseteq\mathcal{X}$ be the initial set, $\mathcal{X}_{U}\subseteq\mathcal{X}$ be the unsafe set, $\mathbb{X}(\mathcal{X}_{I},t,d(t))$ be the set of trajectories initialized in $\mathcal{X}_{I}$. Then the formal definition of the safety property is given as follows.
\begin{definition}[Safety]
\label{thm_def_safety}
Given (\ref{eq_ode}), $\mathcal{X}$, $\mathcal{X}_{I}$, $\mathcal{X}_{U}$ and $\mathcal{D}$, the \emph{safety} property holds if there exists no time instant $T\geq 0$ and no piecewise constant bounded disturbance $d:[0,T]\longrightarrow \mathcal{D}$ such that $\mathbb{X}(\mathcal{X}_{I},t,d(t))\cap \mathcal{X}_{U}\neq\varnothing$ for any $t \in [0,T]$.
\end{definition}\par
\begin{definition}[Region of Safety]
\label{thm_def_ros}
A set that only initializes trajectories with the property in Definition \ref{thm_def_safety} is called a \emph{region of safety}.
\end{definition}\par
Within the bounded set $\mathcal{D}$, the safety property above is defined in the worse-case scenario as well as the region of safety since there is no further limits on disturbance value. Then safety can be verified by the following theorem.
\begin{theorem}
\label{thm_fundamental_barrier}
Let the system $\dot{x}=f(x,d)$, and the sets $\mathcal{X}$, $\mathcal{X}_{I}$, $\mathcal{X}_{U}$ and $\mathcal{D}$ be given, with $f$ continuous. If there exists a differentiable function $B:\mathbb{R}^{n}\longrightarrow \mathbb{R}$ such that
\begin{align}
B(x)\leq 0& \qquad \forall x \in \mathcal{X}_{I}\label{eq_barrier_1}\\
B(x)> 0& \qquad \forall x \in \mathcal{X}_{U}\label{eq_barrier_2}\\
\dfrac{\partial B}{\partial x}f(x,d)<0& \qquad \forall (x,d) \in \mathcal{X}\times\mathcal{D} \quad \mathrm{s.t.} \quad B(x)=0 \label{eq_barrier_3}
\end{align}
then the safety of the system in the sense of Definition \ref{thm_def_safety} is guaranteed \cite{barrier}.
\end{theorem}\par
$B(x)$ is called a barrier certificate. The zero level set of $B(x)$ defines an invariant set containing $\mathcal{X}_{I}$, that is, no trajectory starting in $\mathcal{X}_{I}$ can leave. Thus, $\mathcal{X}_{I}$ is a region of safety (ROS) due to the existence of $B(x)$. Eq. (\ref{eq_barrier_3}) relaxes the passivity condition from the state space to the zero level set of $B(x)$ and thus, reduces conservatism. The other source of conservatism is the initial set $\mathcal{X}_{I}$ usually represented by a ball containing the equilibrium point. However, this set could change under disturbances. Based on this observation, we propose to solve the following problem.
\begin{problem}
\label{thm_max_volume}
Let $\dot{x}=f(x,d)$, $\mathcal{X}$, $\mathcal{X}_{U}$ and $\mathcal{D}$ be given. The region of safety $\mathcal{X}_{I}$ is obtained by solving:
\begin{align*}
&\max_{\mathcal{X}_{I},B(x)} \quad \text{Volume}(\mathcal{X}_{I}) \\
& \text{subject to}\\
& B(x)\leq 0  \quad \forall x \in \mathcal{X}_{I} \\
& B(x)> 0  \quad \forall x \in \mathcal{X}_{U} \\
&\dfrac{\partial B}{\partial x}f(x,d)<0   \quad \forall (x,d) \in \mathcal{X}\times\mathcal{D} \quad \mathrm{s.t.} \quad B(x)=0
\end{align*}
\end{problem}\par
\begin{figure}[h]
	\centering
	\includegraphics[scale=0.4]{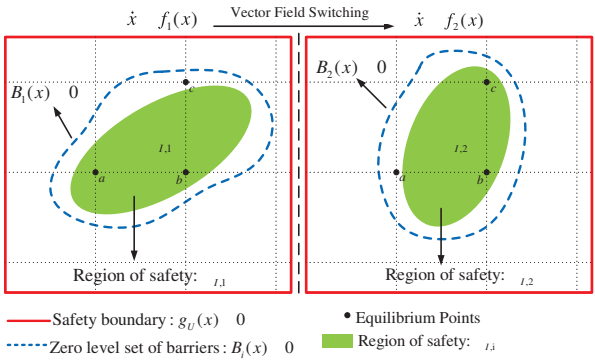}
	\caption{Region of safety and corresponding barriers under different vector fields.}
	\label{fig_ROS_switching_principle}
\end{figure}

\begin{remark}
The importance of introducing the concept of ROS and Problem \ref{thm_max_volume} stems from the fact that we have need to work with the initial sets instead of the invariant sets. Consider the invariant sets $\left\lbrace x \in \mathbb{R}^{n}:B_{i}(x)\leq 0\right\rbrace$ and ROS $\mathcal{X}_{I,i}$ (green regions) with $i=1,2$ calculated under different vector field (or modes) $f_{1}(x)$ and $f_{2}(x)$ with the same safety limits and $\mathcal{D}$ as shown in Fig. \ref{fig_ROS_switching_principle}. Consider set up an emergency alert for $f_{1}(x)$. Based on Theorem \ref{thm_fundamental_barrier} once the trajectory crosses $B_{1}(x)=0$, the alert is triggered. The dynamics after the alert is not our concern (black dash lines in Fig. \ref{fig_WTG_transition}). The safety supervisor of wind turbine shutdown in \cite{shutdown} is based on this principle. Now consider a transition from $f_{1}(x)$ to $f_{2}(x)$, since the state variables are continuous, they will evolve according to $f_{2}(x)$ beginning at the last points before transition. As a result, safety can be guaranteed only if this initial value belongs to the ROS under $f_{2}(x)$. In Fig. \ref{fig_ROS_switching_principle}, only the point $b$ is safe under $f_{1}(x)$ and after switching. The point $a$ is safe under $f_{1}(x)$ but unsafe after the transition to $f_{2}(x)$. The point $c$ is a safe switching point but not safe under $f_{1}(x)$.
\end{remark}
\begin{proposition}
\label{thm_ros_hybrid_principle}
In a hybrid system with several modes, safe switching to mode $i$ is guaranteed if the trajectory of the current mode belongs to the ROS of mode $i$. Moreover, if ROS is represented by some sublevel set of a continuous function in terms of system states, then this function represents a safe switching guard.
\end{proposition}\par
Proposition \ref{thm_ros_hybrid_principle} illustrates the fundamental principle for switching analysis in a hybrid system. Hence, the purpose of Problem \ref{thm_max_volume} is to lessen conservatism. Before we introduce the proposed iterative algorithm to approximate the solution to Problem \ref{thm_max_volume}, the computational techniques for Theorem \ref{thm_fundamental_barrier} under polynomial data is introduced. Polynomial data denotes that all sets are basic closed semi-algebraic sets (hence defined by finitely many polynomial inequalities and equality constraints) and vector fields are polynomial \cite{siam_rev}. Then the property that the polynomials are non-negative on the given semi-algebraic sets can be checked by sums of squares decomposition, which can be further converted to semidefinite programming (SDP) \cite{sos}. From now on all functions are assumed to be polynomial unless specified otherwise. Then the conditions in Theorem \ref{thm_fundamental_barrier} can be written into a sums of squares programming (SOSP) problem. First let us denote by $\varSigma^{2}\left[ x\right] $ the space of SOS polynomials, and by $\varSigma^{2}\left[ x\right]_{p} $ the space of SOS polynomials of degree at most $2p$.
\begin{problem}
\label{thm_sos}
Let $\mathcal{X}=\left\lbrace x \in \mathbb{R}^{n}:g_{X}(x)\geq 0\right\rbrace $, $\mathcal{X}_{I}=\left\lbrace x \in \mathbb{R}^{n}:g_{I}(x)\geq 0\right\rbrace $, $\mathcal{X}_{U}=\left\lbrace x \in \mathbb{R}^{n}:g_{U}(x)\geq 0\right\rbrace $, and $\mathcal{D}=\left\lbrace d \in \mathbb{R}^{m}:g_{D}(d)\geq 0\right\rbrace$, which are represented by the zero superlevel set of the polynomials $g_{X}(x)$, $g_{I}(x)$, $g_{U}(x)$, and $g_{D}(d)$, respectively, and some small positive number $\epsilon$ be given. Then
\begin{align}
-B(x)-\lambda_{I}(x)g_{I}(x)& \in \varSigma^{2}\left[ x\right] \label{eq_sos_1}\\
 B(x)-\epsilon - \lambda_{U}(x)g_{U}(x)& \in \varSigma^{2}\left[ x\right] \label{eq_sos_2}\\
\begin{split}
-\dfrac{\partial B}{\partial x}(x)f(x,d)-\lambda_{D}(x,d)g_{D}(d)\\-\lambda_{X}(x,d)g_{X}(x)-\lambda_{B}(x,d)B(x) &\in \varSigma^{2}\left[ x\right]\label{eq_sos_3}
\end{split}
\end{align}
with multipliers $\lambda_{I}(x)$, $\lambda_{U}(x)$, $\lambda_{X}(x,d)$, $\lambda_{D}(x,d)$ and $\lambda_{B}(x,d)$ SOS polynomials.
\end{problem}\par
Conversion of Problem \ref{thm_sos} to SDP has been implemented in solvers such as SOSTOOLS \cite{sostools} or the sum of squares module \cite{sos_yalmip} in YALMIP \cite{yalmip}. Then the powerful SDP solver MOSEK \cite{mosek} can be employed. Now let us introduce the algorithm to approximate the solution of Problem \ref{thm_max_volume}.
\begin{algorithm}
Let $\mathcal{X}=\left\lbrace x \in \mathbb{R}^{n}:g_{X}(x)\geq 0\right\rbrace $, $\mathcal{X}_{U}=\left\lbrace x \in \mathbb{R}^{n}:g_{U}(x)\geq 0\right\rbrace $, $\mathcal{D}=\left\lbrace d \in \mathbb{R}^{m}:g_{D}(d)\geq 0\right\rbrace$, which are represented by the zero superlevel of the polynomials $g_{X}(x)$, $g_{U}(x)$ and $g_{D}(d)$, respectively, some small positive number $\epsilon$, initial order $2p$ and maximal order $2p_{\text{max}}$ for barrier certificate computation be given.
\begin{itemize}
\item \textbf{Initialization} Let $x_{0}^{i}$ for $i=1,\cdots,N$ be several initial points with safety verified, and $\mathcal{X}_{I,i}=\left\lbrace x \in \mathbb{R}^{n}:g_{I,i}(x)\geq 0\right\rbrace $ represent a small ball centered at $x_{0}^{i}$. Choose $\lambda_{B}(x,d)$ equal to a sufficiently small positive real number $r$ and solve the following SOS optimization for $i=1,\cdots,N$:
\begin{align*}
-B^{(0)}(x)-\lambda^{(0)}_{I,i}(x)g_{I,i}(x)& \in \varSigma^{2}\left[ x\right]  \\
B^{(0)}(x)-\epsilon - \lambda^{(0)}_{U}(x)g_{U}(x)& \in \varSigma^{2}\left[ x\right] \\
\begin{split}
-\dfrac{\partial B^{(0)}}{\partial x}(x)f(x,d)-\lambda^{(0)}_{D}(x,d)g_{D}(d)\\-\lambda^{(0)}_{X}(x,d)g_{X}(x)-rB^{(0)}(x) &\in \varSigma^{2}\left[ x\right]
\end{split}
\end{align*}
\item \textbf{Iteration \emph{k}}
\begin{itemize}
\item[(a)] Fix the barrier certificate $B^{(k-1)}(x)$ from $k-1$ step, solve the SOS optimization for multiplier $\lambda^{(k_{a})}_{B}(x,d)$:
\begin{align*}
\begin{split}
-\dfrac{\partial B^{(k-1)}}{\partial x}(x)f(x,d)-\lambda^{(k_{a})}_{D}(x,d)g_{D}(d)\\-\lambda^{(k_{a})}_{X}(x,d)g_{X}(x)-\lambda^{(k_{a})}_{B}(x,d)B^{(k-1)}(x) &\in \varSigma^{2}\left[ x\right]
\end{split}
\end{align*}
\item[(b)] Fix the barrier certificate $B^{(k-1)}(x)$ from $k-1$ step, the multiplier $\lambda^{(k_{a})}_{B}(x,d)$ from $k$ (a) step, solve the following SOS optimization for $B^{(k)}(x)$:
\begin{align*}
-B^{(k)}(x)-\lambda^{(k)}_{I}(x)B^{(k-1)}(x)& \in \varSigma^{2}\left[ x\right]\\
B^{(k)}(x)-\epsilon - \lambda^{(k)}_{U}(x)g_{U}(x)& \in \varSigma^{2}\left[ x\right] \\
\begin{split}
-\dfrac{\partial B^{(k)}}{\partial x}(x)f(x,d)-\lambda^{(k)}_{D}(x,d)g_{D}(d)\\-\lambda^{(k)}_{X}(x,d)g_{X}(x)-\lambda^{(k_{a})}_{B}(x,d)B^{(k)}(x) &\in \varSigma^{2}\left[ x\right]
\end{split}
\end{align*}
\item[(c)] If step $k$ (b) is feasible, then let $k=k+1$. If infeasible, then increase the polynomial order of $B^{(k)}$ by two, i.e., $2p=2p+2$. If $p=p_{\text{max}}$ but step $k$ (b) is still infeasible, then the algorithm stops and $\mathcal{X}^{*}_{I}=\left\lbrace x:B^{(k-2)}(x)\leq 0\right\rbrace $ with $B^{(k-1)}(x)$ the barrier.
\end{itemize}
\end{itemize}
\end{algorithm}\par
The key idea of the proposed algorithm is to use the zero level set of a feasible barrier certificate as an initial condition and to search for a larger invariant set. Once feasible, this initial condition becomes the ROS due to the existence of corresponding invariant sets. A judicious choice of the initial points in the initialization step can reduce the number of iterations, and also helps to have a precise estimate in certain sub-dimensions, if a full dimensional estimate is hard due to computational complexity.\par

\section{Hybrid Reduced-order Model of WTG via SMA-based Model Reduction}
\begin{figure}[h]
	\centering
	\includegraphics[scale=0.35]{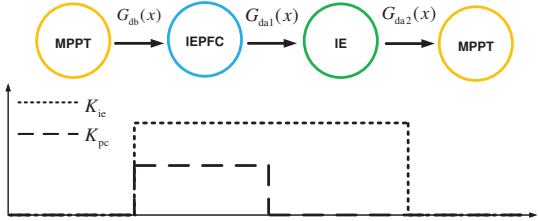}
	\caption{Modelling deadband as a hybrid transition system.}
	\label{fig_deadband_hybrid_model}
\end{figure}

Consider the active power control in Fig. \ref{fig_WTG_control}. The WTG is assumed to operate at partial loaded condition, which means $P_{\text{gen}}<P_{\text{gen}}^{\text{max}}$ and $\omega_{r}^{\text{min}}<\omega_{r}<\omega_{r}^{\text{max}}$. And no pitch angle control is considered. The control signal sent to the voltage-source converter (VSC) is given by
\begin{equation}
\label{eq_Pref}
P_{\text{ref}}=C_{\text{opt}}\omega^{3}_{r}+K_{\text{ie}}(\omega_{\text{grid}})\Delta\dot{\omega}+K_{\text{pc}}(\omega_{\text{grid}})\Delta\omega
\end{equation}
where $C_{\text{opt}}\omega^{3}_{r}$, $K_{\text{ie}}(\omega_{\text{grid}})\Delta\dot{\omega}$, and $K_{\text{pc}}(\omega_{\text{grid}})\Delta\omega$ provide functionalities of maximum power point tracking (MPPT), inertia emulation (IE) and primary frequency control (PFC), respectively. $K_{\text{ie}}$ and $K_{\text{pc}}$ are equal to zero when  $\mid\Delta\omega_{\text{grid}}\mid<\Delta\omega_{\text{deadband}}$, or equal to certain pre-set values otherwise. The deadband effect is equivalent to the switching guard $G_{\text{db}}(x)$ in the hybrid system illustrated in Fig. \ref{fig_deadband_hybrid_model}. $G_{\text{da}1}(x)$ and $G_{\text{da}2}(x)$ represent signals (dash line in Fig. \ref{fig_WTG_control}) for deactivating the corresponding support in order to have a faster frequency restoration after safety is preserved. It is worth mentioning that if ROSs of corresponding modes are employed for those guards, then by Proposition \ref{thm_ros_hybrid_principle}, safety can be guaranteed.
\subsection{Model Reduction}
\label{sec_model}
As shown in \cite{ROAviaSOS}, there is a trade-off between system dimension and the order of polynomials for set representation. Since the active power variations and frequency dynamics in power system are dominantly governed by mechanical dynamics and modes, a reduced-order system is desired in analysis so that higher order polynomials can be used for better estimation of the ROS. The selective modal analysis (SMA) based model reduction has proven to be successful in capturing active power dynamics of WTG \cite{hector} and is chosen for our study.\par
Consider a type-3 WTG connected to a reference bus. The detailed model and the meaning of each variable is given in Appendix \ref{appendix_all_model}. Linearize (\ref{eq_turbine}), (\ref{eq_WTG_diff}), (\ref{eq_WTG_alg}) and (\ref{eq_network_alg}) about the equilibrium point given in Appendix \ref{appendix_data}. Keep $\Delta\dot{\omega}$ and $\Delta\omega$ as input variables and omit the variations of $v_{\text{wind}}$. Then the following model is obtained:
\begin{subequations}
\label{eq_linear_overall}
\begin{align}
\begin{split}
\left[\begin{array}{c}\Delta\dot{x}\\0 \end{array}\right] &=\left[ \begin{array}{cc} 
A_{s} & B_{s}\\
C_{s} & D_{s}
\end{array} 
\right]\left[\begin{array}{c}\Delta x\\\Delta y \end{array} \right] 
 + \left[\begin{array}{c} M_{s1}\\N_{s1} \end{array} \right]\Delta\dot{\omega}\\
& + \left[\begin{array}{c} M_{s2}\\N_{s2} \end{array} \right]\Delta\omega
\end{split}\\
\Delta P_{\text{gen}}&=\left[E_{s} \quad F_{s} \right]\left[\begin{array}{c}\Delta x\\\Delta y \end{array} \right]
\end{align}
\end{subequations}
where
\begin{align}
x&=\left[E_{qD},E_{dD},\omega_{r},x_{1},x_{2},x_{3},x_{4}\right]^{T}\\
y&=\left[V_{qr},V_{dr},I_{qr},I_{dr},P_{\text{gen}},Q_{\text{gen}},I_{ds},I_{qs},V_{D},\theta_{D}\right]^{T}
\end{align}
Using Kron reduction \cite{kundur1994power} on Eq. (\ref{eq_linear_overall}) yields the following state-space model:
\begin{subequations}
\label{eq_linear_ss_full}
\begin{align}
\Delta\dot{x}&=A_{\text{sys}}\Delta x + B_{\text{sys}1}\Delta\dot{\omega}+B_{\text{sys}2}\Delta\omega\\
\Delta P_{\text{gen}}&=C_{\text{sys}}\Delta x +  D_{\text{sys}1}\Delta\dot{\omega}+D_{\text{sys}2}\Delta\omega
\end{align}
\end{subequations}
where
\begin{align*}
 A_{\text{sys}}&=A_{s}-B_{s}D^{-1}_{s}C_{s} &  C_{\text{sys}}&=E_{s}-F_{s}D^{-1}_{s}C_{s} \\
  B_{\text{sys}1}&=M_{s1}-B_{s}D^{-1}_{s}N_{s1} & D_{\text{sys}1}&=-F_{s}D^{-1}_{s}N_{s1} \\
  B_{\text{sys}2}&=M_{s2}-B_{s}D^{-1}_{s}N_{s2} & D_{\text{sys}2}&=-F_{s}D^{-1}_{s}N_{s2}
\end{align*}\par
The WTG rotor speed $\Delta\omega_{r}$ dynamic is closely related to its active power output, and the mode where $\Delta\omega_{r}$ has the highest participation would capture the relevant active power dynamics. Therefore, $\Delta\omega_{r}$ is considered as the most relevant state, and the other states are less relevant and denoted as $z(t)$. Eq. (\ref{eq_linear_ss_full}) can be rearranged as
\begin{subequations}
\label{eq_linear_ss_arrange}
\begin{align}
\begin{split}
\left[\begin{array}{c}\Delta\dot{\omega_{r}}\\\dot{z} \end{array}\right] &=\left[ \begin{array}{cc} 
A_{11} & A_{12}\\
A_{21} & A_{22}
\end{array} 
\right]\left[\begin{array}{c}\Delta\omega_{r}\\z \end{array} \right] \\
& + \left[\begin{array}{c} B_{r1}\\B_{z1} \end{array} \right]\Delta\dot{\omega}
 + \left[\begin{array}{c} B_{r2}\\B_{z2} \end{array} \right]\Delta\omega
\end{split}\\
\begin{split}
\Delta P_{\text{gen}}&=\left[C_{r} \quad C_{z} \right] \left[\begin{array}{c}\Delta\omega_{r}\\z \end{array}\right]\\
&+D_{\text{sys}1}\Delta\dot{\omega} +D_{\text{sys}2}\Delta\omega
\end{split}
\end{align}
\end{subequations}
The less relevant dynamics are:
\begin{equation}
\label{eq_linear_ss_less}
\dot{z}=A_{22}z+A_{21}\Delta\omega_{r} + B_{z1}\Delta\dot{\omega} + B_{z2}\Delta\omega
\end{equation}
Thus, the most relevant dynamic is described by: \\
\begin{equation}
\label{eq_linear_ss_more}
\Delta\dot{\omega}_{r}=A_{11}\Delta\omega_{r}+A_{12}z + B_{r1}\Delta\dot{\omega} + B_{r2}\Delta\omega
\end{equation}
In (\ref{eq_linear_ss_more}), $z$ can be represented by the following expression:
\begin{equation}
\label{eq_sol_z}
\begin{split}
z(t)&=\underbrace{e^{A_{22}(t-t_{0})}z(t_{0})+\int_{t_{0}}^{t}e^{A_{22}(t-\tau)}A_{21}\Delta\omega_{r}(\tau)d\tau}_{\text{response without control input}}\\
&+\underbrace{\int_{t_{0}}^{t}e^{A_{22}(t-\tau)}B_{z1}\Delta\dot{\omega}(\tau)d\tau}_{\text{response under inertia emulation}}\\
&+\underbrace{\int_{t_{0}}^{t}e^{A_{22}(t-\tau)}B_{z2}\Delta\omega(\tau)d\tau}_{\text{response under primary frequency control}}
\end{split}
\end{equation}
Using the most relevant mode, $\Delta\omega_{r}(\tau)$ can be expressed as \cite{hector}: 
\begin{equation}
\label{eq_sol_r}
\Delta\omega_{r}(\tau)=c_{r}v_{r}e^{\lambda_{r}\tau}
\end{equation}
where $\lambda_{r}$ is the relevant eigenvalue, $v_{r}$ is the corresponding eigenvector and $c_{r}$ is an arbitrary constant. The accuracy of (\ref{eq_sol_r}) is guaranteed by the dominant term of $\Delta\omega_{r}$, which can be used in solving the first integral in (\ref{eq_sol_z}). Since $A_{22}$ is Hurwitz and its largest eigenvalue is much smaller than $\lambda_{r}$, the natural response will decay faster and can be omitted. The essential reason is that $A_{22}$ represents electrical dynamics which are faster than the electro-mechanical dynamic represented by $\lambda_{r}$. Then the response without control input in (\ref{eq_sol_z}) will approximately equal to the forced response represented as follows:
\begin{align}
\label{eq_sol_integral_1}
\underbrace{e^{A_{22}(t-t_{0})}z(t_{0})+\int_{t_{0}}^{t}e^{A_{22}(t-\tau)}A_{21}\Delta\omega_{r}(\tau)d\tau}_{\text{response without control input}}\\
\approx (\lambda_{r}I-A_{22})^{-1}A_{21}\Delta\omega_{r}
\end{align}
The rate of change of frequency (RoCoF) $\Delta\dot{\omega}$ and the stabilized frequency deviation $\Delta\omega$ are assumed to be fixed during the time window of interest, then the other two integrals are easily calculated as
\begin{align}
\label{eq_sol_integral_rest}
&\underbrace{\int_{t_{0}}^{t}e^{A_{22}(t-\tau)}B_{z1}\Delta\dot{\omega}(\tau)d\tau}_{\text{response under inertia emulation}}\approx (-A_{22})^{-1}B_{z1}\Delta\dot{\omega}\\
&\underbrace{\int_{t_{0}}^{t}e^{A_{22}(t-\tau)}B_{z2}\Delta\omega(\tau)d\tau}_{\text{response under primary frequency control}}\approx (-A_{22})^{-1}B_{z2}\Delta{\omega}
\end{align}
Finally, the reduced-order WTG model with control inputs is
\begin{subequations}
\label{eq_linear_ss_reduced}
\begin{align}
\Delta\dot{\omega}_{r}&=A_{\text{rd}}\Delta\omega_{r} + B_{\text{rd}1}\Delta\dot{\omega}  + B_{\text{rd}2}\Delta\omega\\
\Delta P_{\text{gen}}&=C_{\text{rd}}\Delta\omega_{r} + D_{\text{rd}1}\Delta\dot{\omega}  + D_{\text{rd}2}\Delta\omega 
\end{align}
\end{subequations}
where
\begin{align*}
& A_{\text{rd}}=A_{11}+A_{12}(\lambda_{r}I-A_{22})^{-1}A_{21} \\
& C_{\text{rd}}=C_{r}+C_{z}(\lambda_{r}I-A_{22})^{-1}A_{21}\\
& B_{\text{rd}1}=B_{r1} + A_{12}(-A_{22})^{-1}B_{z1}\\
& D_{\text{rd}1}=D_{\text{sys}1} + C_{z}(-A_{22})^{-1}B_{z1}\\
& B_{\text{rd}2}=B_{r2} + A_{12}(-A_{22})^{-1}B_{z2}\\
& D_{\text{rd}2}=D_{\text{sys}2} + C_{z}(-A_{22})^{-1}B_{z2}
\end{align*}
\subsection{Quantification of Frequency Support from WTG}
\label{subsec_quan}
Consider the swing equation with the active power increment $\Delta P_{\text{gen}}$ from WTGs
\begin{equation}
\label{eq_swing_standard}
\Delta\dot{\omega}=\frac{\omega_{s}}{2H}(\Delta P_{m}+\Delta P_{\text{gen}}-\Delta P_{e}-\dfrac{D}{\omega_{s}}\Delta\omega)
\end{equation}
The WTG active power output in (\ref{eq_swing_standard}) due to the signal $\Delta\dot{\omega}$ and $\Delta{\omega}$ will influence the values of $H$ and $D$ independently.\par
To evaluate the emulated inertia, the terms related to the PFC, i.e., $B_{\text{rd}2}$ and $D_{\text{rd}2}$ in (\ref{eq_linear_ss_reduced}), are set to zero. The explicit forced output response of (\ref{eq_linear_ss_reduced}) due to $\Delta\dot{\omega}$ is given by
\begin{equation}
\label{eq_output_solution_dfdt}
\Delta P_{\text{gen}}(t)=C_{\text{rd}}\int_{t_{0}}^{t}e^{A_{\text{rd}}(t-\tau)}B_{\text{rd}1}\Delta\dot{\omega}(\tau)d\tau+D_{\text{rd}1}\Delta\dot{\omega}(t)
\end{equation}
During the time window of inertia response $T_{h}=\left\lbrace t:0\leq t \leq t_{h} \right\rbrace $, the RoCoF is approximately fixed. Then $\Delta\dot{\omega}$ can be pulled outside the integral and integrating (\ref{eq_output_solution_dfdt}) with $t_{0}=0$ yields
\begin{equation}
\label{eq_output_ie}
\Delta P_{\text{gen}}(t)=(D_{\text{rd}1}-C_{\text{rd}}A_{\text{rd}}^{-1}(I-e^{A_{\text{rd}}t})B_{\text{rd}1})\Delta\dot{\omega}
\end{equation}
Substituting (\ref{eq_output_ie}) back into (\ref{eq_swing_standard}) and rearranging the state yields
\begin{equation}
\label{eq_swing_ie}
\Delta\dot{\omega}=\frac{\omega_{s}}{2H+2H_{e}(t)}(\Delta P_{m}-\Delta P_{e}-\dfrac{D}{\omega_{s}}\Delta\omega)
\end{equation}
where
\begin{equation}
\label{eq_He}
H_{e}(t)=0.5\omega_{s}(-D_{\text{rd}1}+C_{\text{rd}}A_{\text{rd}}^{-1}(I-e^{A_{\text{rd}}t})B_{\text{rd}1})
\end{equation}\par
To evaluate the emulated load-damping effect, the terms related to inertia emulation, i.e., $B_{\text{rd}1}$ and $D_{\text{rd}1}$ in (\ref{eq_linear_ss_reduced}), are set to zero. The explicit forced output response of (\ref{eq_linear_ss_reduced}) due to $\Delta\omega$ is given as
\begin{equation}
\label{eq_output_solution_df}
\Delta P_{\text{gen}}(t)=C_{\text{rd}}\int_{t_{0}}^{t}e^{A_{\text{rd}}(t-\tau)}B_{\text{rd}2}\Delta\omega(\tau)d\tau+D_{\text{rd}2}\Delta\omega(t)
\end{equation}
After the frequency is stabilized by the governor, i.e., $t\in T_{p}=\left\lbrace t:t_{p}\leq t \leq t_{s} \right\rbrace $, the term $\Delta\omega$ can be pulled outside of the integral and integrating (\ref{eq_output_pfc}) with $t_{0}=t_{p}$ yields
\begin{equation}
\label{eq_output_pfc}
\Delta P_{\text{gen}}(t)=(D_{\text{rd}2}-C_{\text{rd}}A_{\text{rd}}^{-1}(I-e^{A_{\text{rd}}(t-t_{p})})B_{\text{rd}2})\Delta\omega
\end{equation}
Substituting (\ref{eq_output_pfc}) into (\ref{eq_swing_standard}) yields
\begin{equation}
\label{eq_swing_pfc}
\Delta\dot{\omega}=\frac{\omega_{s}}{2H}(\Delta P_{m}-\Delta P_{e}-\dfrac{(D+D_{e}(t))}{\omega_{s}}\Delta\omega)
\end{equation}
where
\begin{equation}
\label{eq_De}
D_{e}(t)=\omega_{s}(-D_{\text{rd}2}+C_{\text{rd}}A_{\text{rd}}^{-1}(I-e^{A_{\text{rd}}(t-t_{p})})B_{\text{rd}2})
\end{equation}
Eq. (\ref{eq_swing_pfc}) and (\ref{eq_De}) have clearly illustrated that PFC is actually emulating load-damping characteristic.
\begin{remark}
The controller gain is preserved under the approximations that the RoCoF $\Delta\dot{\omega}$ and the stabilized frequency deviation $\Delta\omega$ is constant within the time window of interest. The comparison to the full order model will show that this approximation is accurate within these windows. Then the emulated inertia $H_{e}$ and load-damping coefficient $D_{e}$ are expressed in term of $B_{\text{rd}1}$, $D_{\text{rd}1}$ and $B_{\text{rd}2}$, $D_{\text{rd}2}$, which correspond to $K_{\text{ie}}$ and $K_{\text{pc}}$. Note that $H_{e}(t)$ and $D_{e}(t)$ are time-varying and their values are considered to be accurate only within the corresponding time windows, i.e., $T_{h}$ and $T_{p}$.
\end{remark}

\section{Case Study}
\label{sec_case_study}
\begin{figure*}[htbp!]
	\centering
	\begin{minipage}[]{0.24\textwidth}
		\centering
		\includegraphics[scale=0.15]{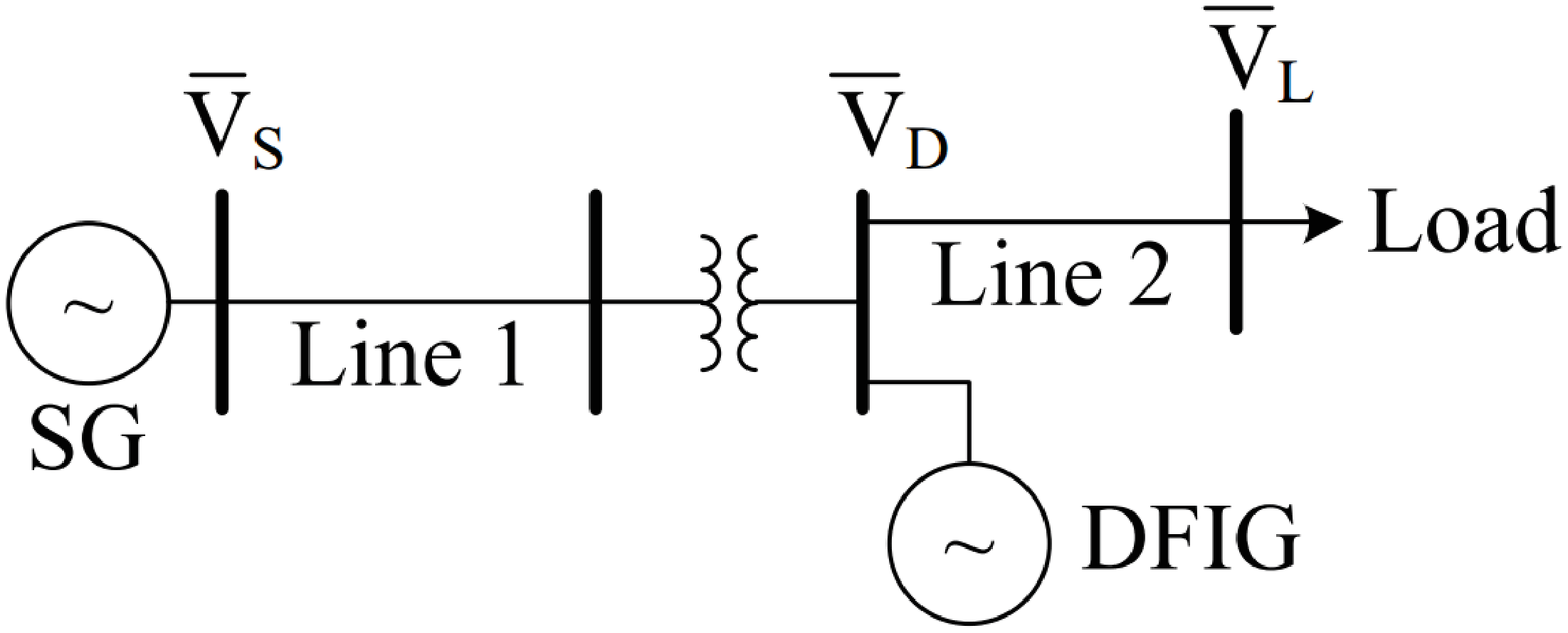}  %  Four_Bus.png
		\caption{Case study: four-bus system \cite{hector}.}
		\label{fig_four_bus}
	\end{minipage}\hfill
	\begin{minipage}[]{0.24\textwidth}
		\centering
		\includegraphics[scale=0.3]{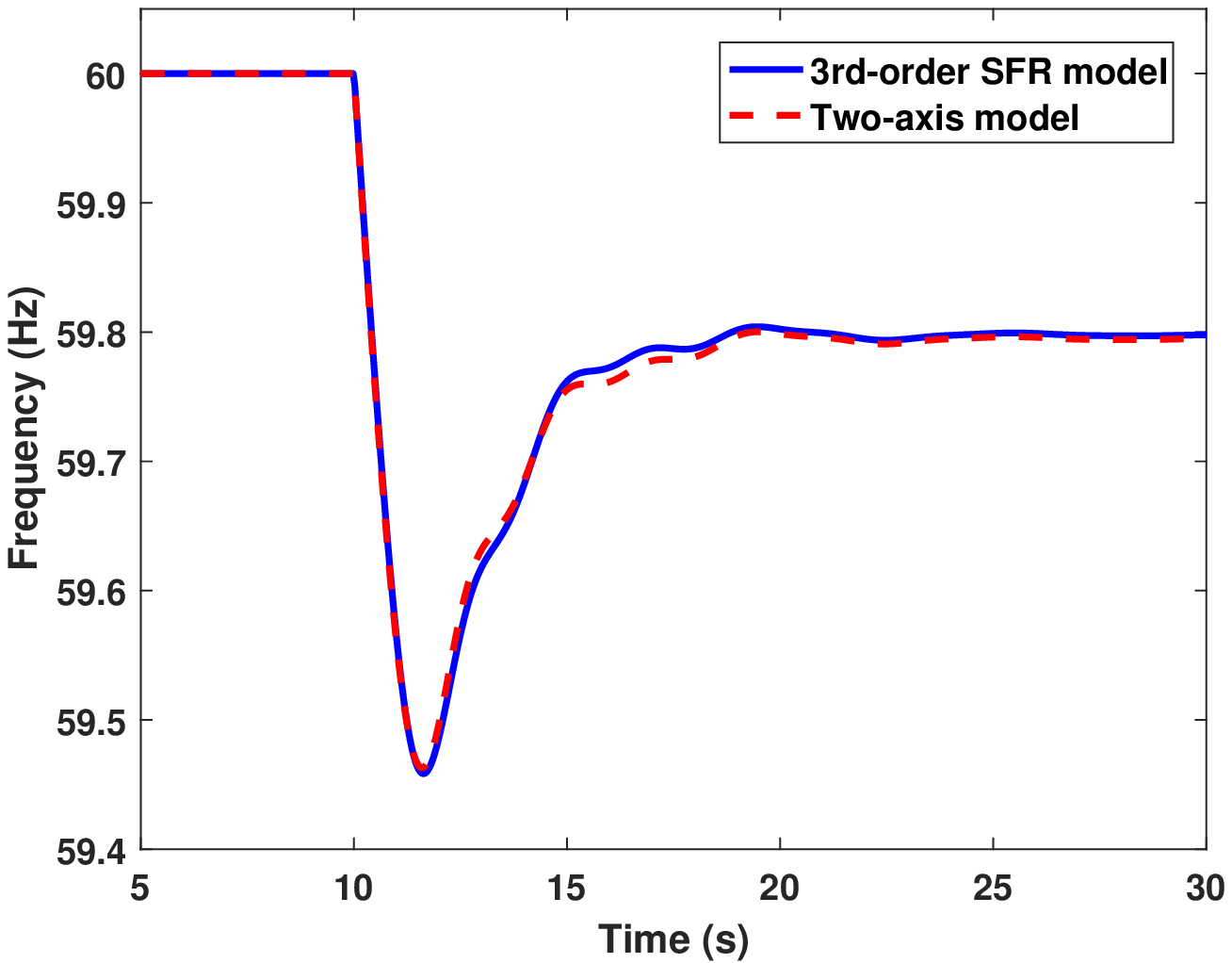}  %SFR.eps
		\caption{Response comparison of system frequency response model and two-axis model.}
		\label{fig_SFR}
	\end{minipage}\hfill
		\begin{minipage}[]{0.24\textwidth}
			\centering
			\includegraphics[scale=0.3]{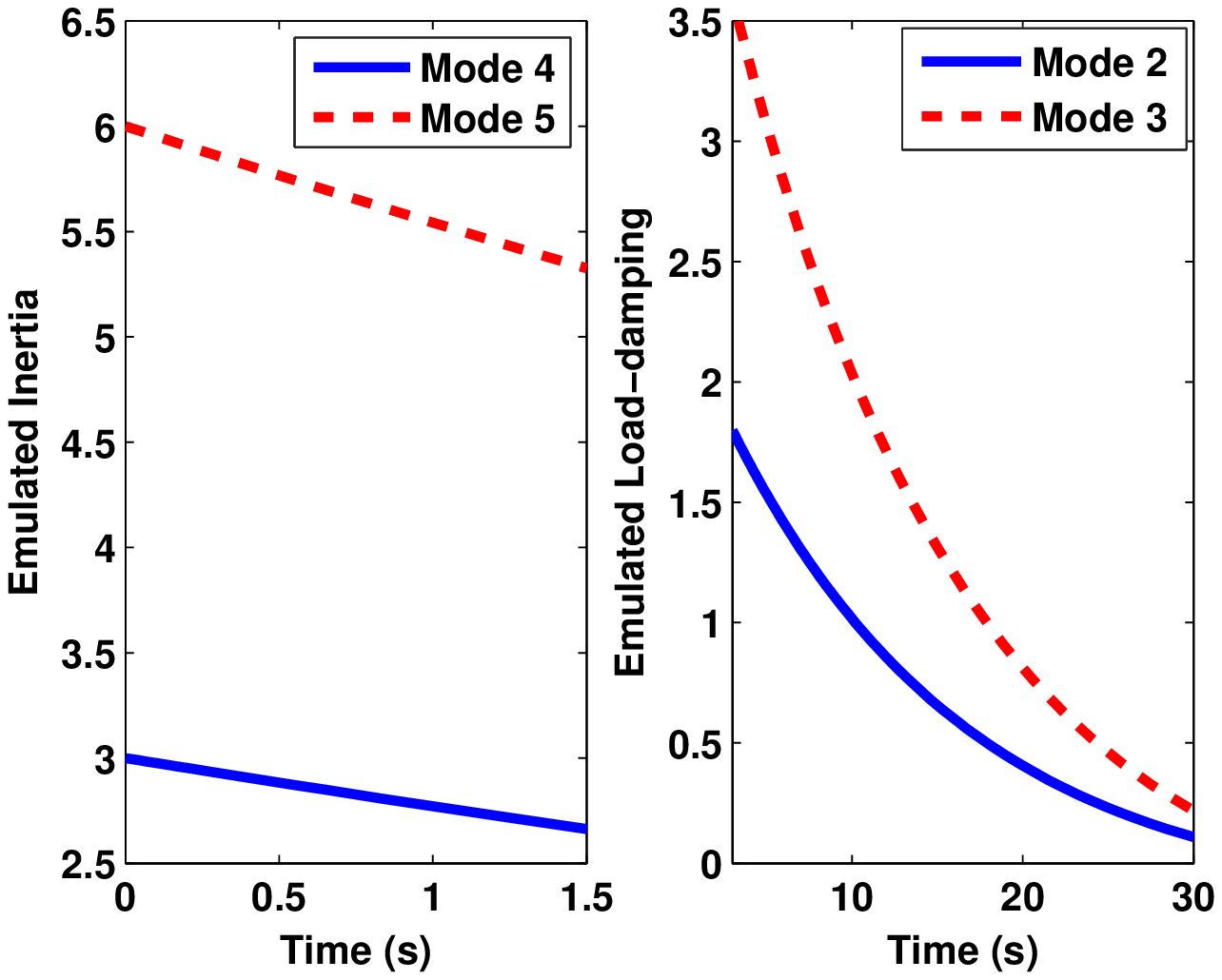}
			\caption{Time-varying emulated inertia and load-damping coefficient.}
			\label{fig_emulated_value}
		\end{minipage}\hfill
				\begin{minipage}[]{0.24\textwidth}
			\centering
			\includegraphics[scale=0.3]{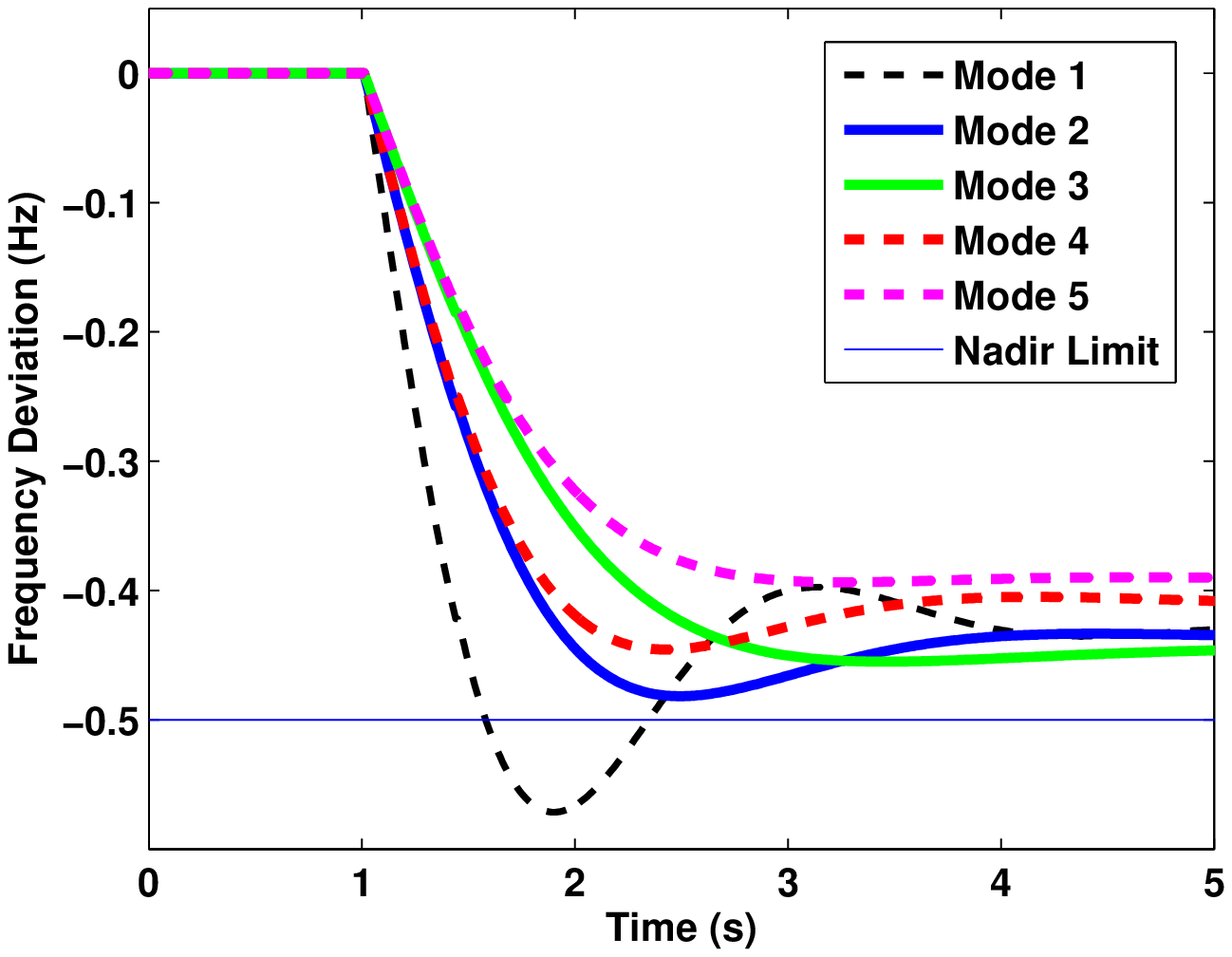}
			\caption{Frequency response of different modes under the worst-case scenario: 150 MW generation loss.}
			\label{fig_FD_Modes}
				\end{minipage}\hfill
\end{figure*}

Consider the four-bus system with a 600 MW thermal plant made up of four identical units in Fig. \ref{fig_four_bus}. The frequency dynamics of the above system can be represented by the classic system frequency response (SFR) model \cite{LFC} as follows:
\begin{subequations}
\label{eq_LFC}
\begin{align}
&\Delta\dot{\omega} =\frac{\omega_{s}}{2H}(\Delta P_{m}-\Delta P_{e}-\dfrac{D}{\omega_{s}}\Delta\omega)\\
&\Delta \dot{P}_{m} =\frac{1}{\tau_{\text{ch}}}(\Delta P_{v}-\Delta P_{m})\\
&\Delta \dot{P}_{v} =\frac{1}{\tau_{g}}(-\Delta P_{v}-\dfrac{1}{R}\Delta\omega)
\end{align}
\end{subequations}
The power flow equation is $\Delta P_{e}=\Delta P_{d}-\Delta P_{\text{gen}}$, where $\Delta P_{d}$ denotes a large disturbance, such as, generation loss or abrupt load changes, and $\Delta P_{\text{gen}}$ given in (\ref{eq_linear_ss_reduced}b) represents the active power variation due to the frequency control loop. Although a one-area case is studied, the SFR model has the potential to describe system frequency response in a complex power network as shown in many recent studies  \cite{GE_SFR} \cite{NREL_GM} \cite{NREL_J}. In this one-area system, as shown in Fig. \ref{fig_SFR}, the response of the SFR model and two-axis nonlinear model are the same under the same disturbance.
 \par
The wind farm is assumed to be an aggregation of 200 individual GE 1.5 MW WTGs with rated speed of 450 rad/s (or 72 Hz) and rated output of 300 MW. Under the operating condition given in Appendix \ref{appendix_data}, the reduced-order WTG model can be obtained with $A_{\text{rd}}=-0.0723$ and $C_{\text{rd}}=0.0127$. $B_{\text{rd}1}$, $D_{\text{rd}1}$ and $B_{\text{rd}2}$, $D_{\text{rd}2}$ with the corresponding $K_{\text{ie}}$ and $K_{\text{pc}}$ are listed in Table \ref{tab_gain_He_De}. The emulated $H_{e}$ and $D_{e}$ are time-varying and shown in Fig. \ref{fig_emulated_value}.\par
The worst-case scenario is assumed to be the loss of one unit (150 MW), which occurs at 1 s. The safety limit is set to be a 0.5 Hz deviation \cite{ORNL} to avoid triggering load shedding \cite{load_shed}. The frequency response of all modes under this scenario is given in Fig. \ref{fig_FD_Modes}. The inertia emulation effect can be observed as the RoCoF becomes slower from the response of Modes 1-3. The ROS (safety switching guard) is calculated under the reduced-order model in Eq. (\ref{eq_LFC}), but the full-order linearized model in Fig. \ref{fig_four_bus} is used for verification. Denote $ x_{\text{rd}}=\left[\Delta\omega,\Delta P_{m},\Delta P_{v},\Delta\omega_{r} \right]$ and $x=\left[\Delta\omega,\Delta P_{m},\Delta P_{v},\Delta E^{\prime}_{qD},\Delta E^{\prime}_{dD},\Delta\omega_{r},\Delta x_{1},\Delta x_{2},\Delta x_{3},\Delta x_{4}\right]$ for theoretical analysis and simulation verification, respectively.
\begin{table}[htbp!]
	\caption{Gain of Frequency Support Mode and Corresponding Matrix Value}
	\centering
	\begin{tabular}{lclclclclclclclcl}
		\toprule 
		Mode & Number & $K_{\text{ie}}$ & $B_{\text{rd}1}$ & $D_{\text{rd}1}$ & $K_{\text{pc}}$ & $B_{\text{rd}2}$ & $D_{\text{rd}2}$ \\
		\midrule
		MPPT & 1 & 0 & 0 & 0 & 0 & 0 & 0\\
		IE & 2 & -0.10 & 0.6246 & -0.10 & 0 & 0 & 0  \\
		IE & 3 &  -0.20 & 1.2492 & -0.20 & 0 & 0 & 0  \\
		IEPFC & 4 & -0.10 & 0.6246 & -0.10 & -0.03 & 0.1874 & -0.03 \\
		IEPFC & 5  & -0.20 & 1.2492 & -0.20 & -0.06 & 0.3748 & -0.06 \\
		\bottomrule
	\end{tabular}
	\label{tab_gain_He_De}
\end{table}\par
\subsection{Model and Algorithm Validation}
\begin{figure*}[htbp!]
	\centering
	\subfloat[]{
		\begin{minipage}[]{0.23\textwidth}
			\centering
			\includegraphics[scale=0.32]{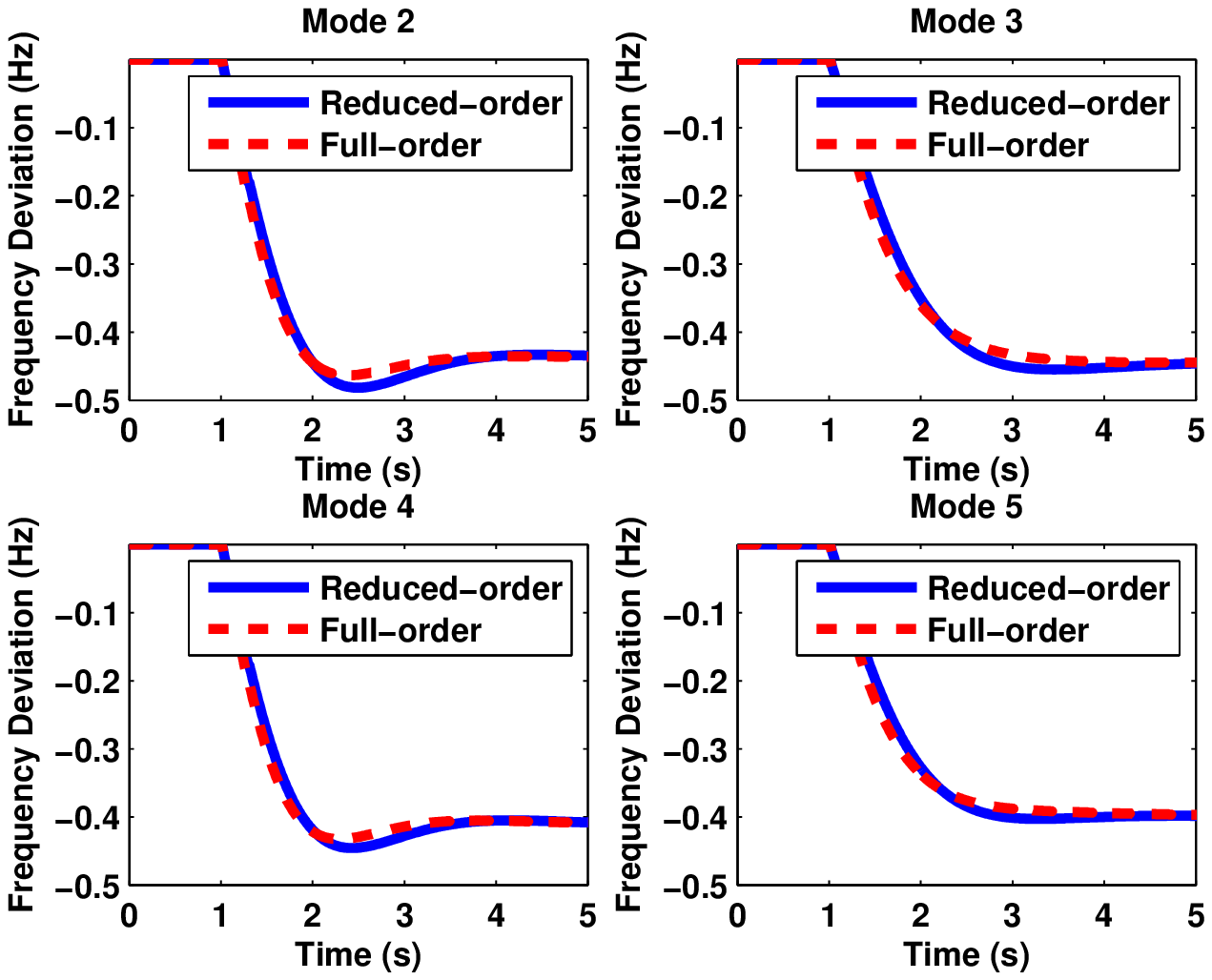}
		\end{minipage}
	}
	\subfloat[]{
		\begin{minipage}[]{0.23\textwidth}
			\centering
			\includegraphics[scale=0.32]{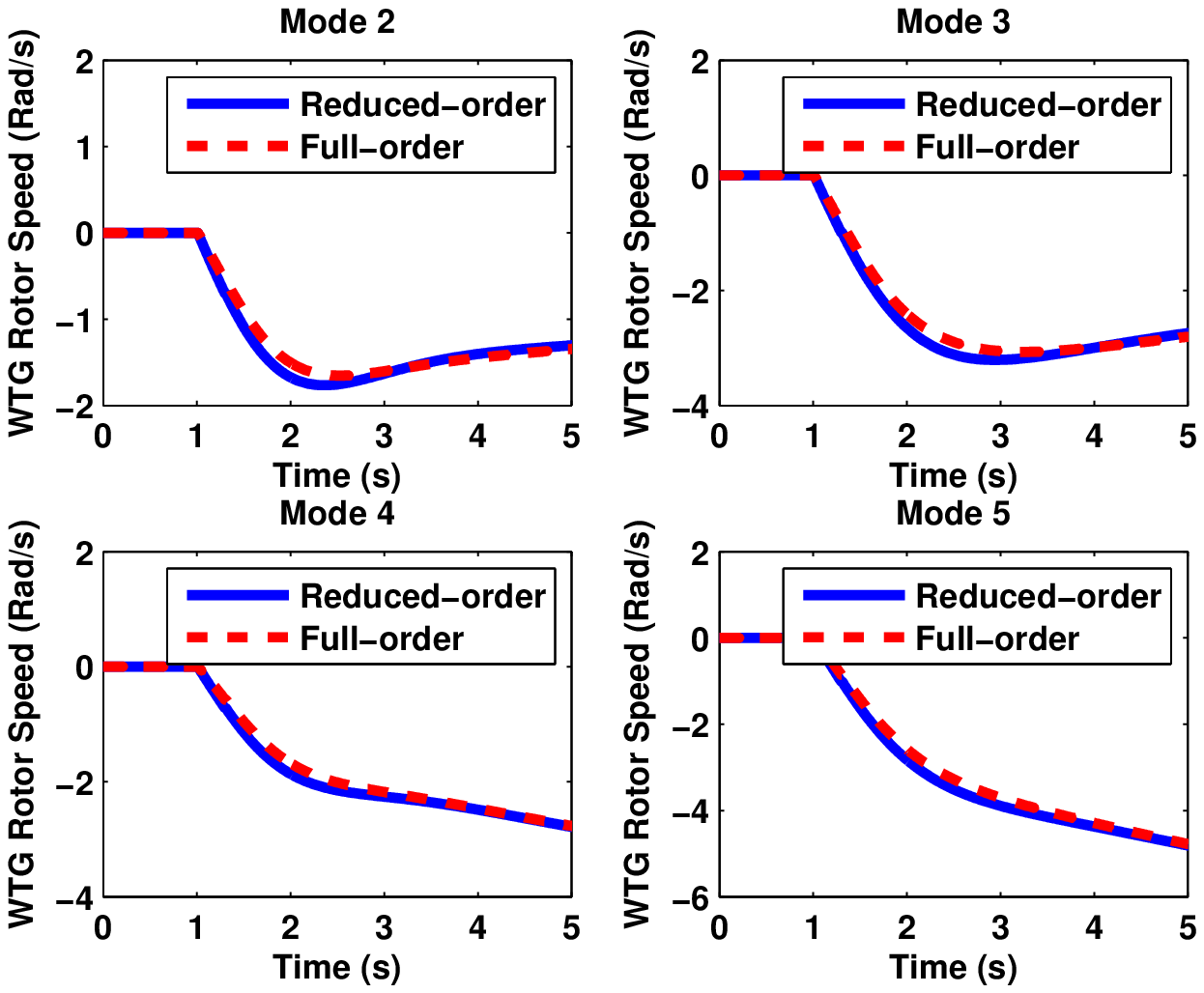}
		\end{minipage}
	}
	\subfloat[]{
		\begin{minipage}[]{0.23\textwidth}
			\centering
			\includegraphics[scale=0.32]{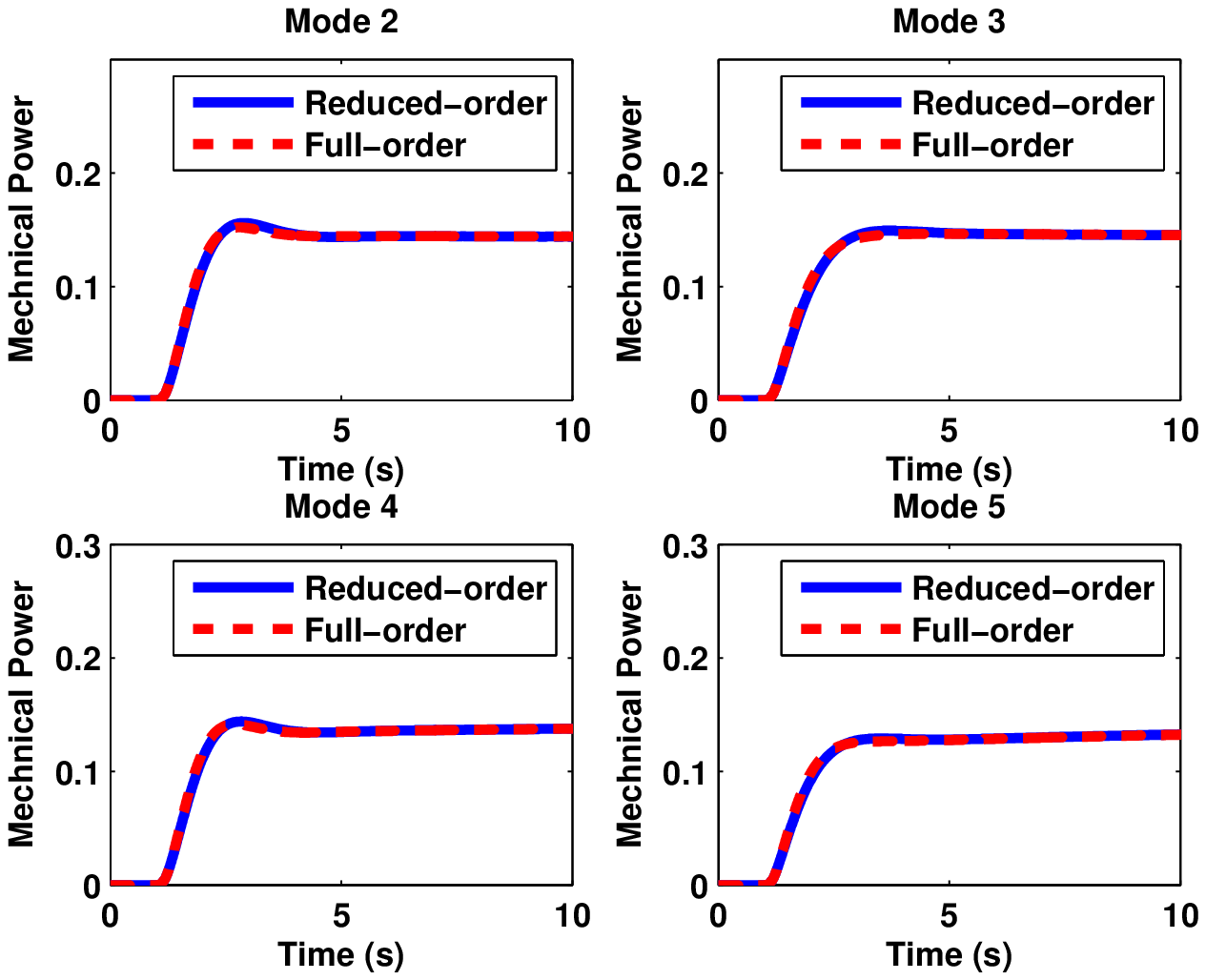}
		\end{minipage}
	}
	\subfloat[]{
		\begin{minipage}[]{0.23\textwidth}
			\centering
			\includegraphics[scale=0.32]{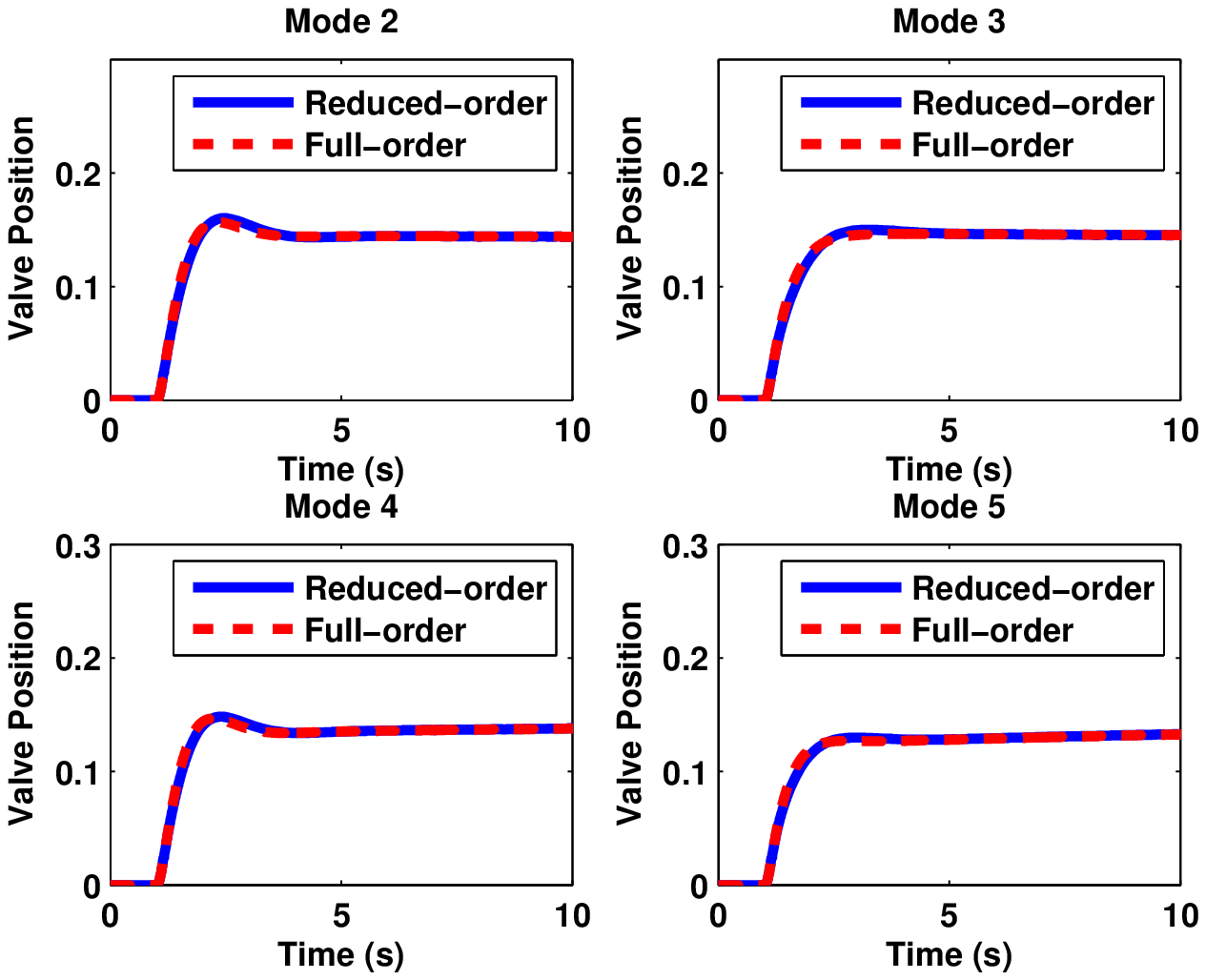}
		\end{minipage}
	}
	\caption{Dynamics between full-order and reduced-order model: (a) Frequency deviation; (b) WTG rotor speed; (c) Turbine-governor mechanical power; (d) Turbine-governor valve position.}
	\label{fig_model_validate}
\end{figure*}

To validate the reduced-order model, consider the worst-case scenario above. The four state variables $\Delta\omega$, $\Delta P_{m}$, $\Delta P_{v}$, $\Delta\omega_{r}$ between reduced-order and full-order model of Mode 2-5 in Table \ref{tab_gain_He_De} are compared in Fig. \ref{fig_model_validate}. The excellent agreement in mode behaviour ensures the reduced-order model based ROS should be sufficient to find the switching for the full-order dynamics.\par
\begin{figure}[h]
	\centering
	\includegraphics[scale=0.4]{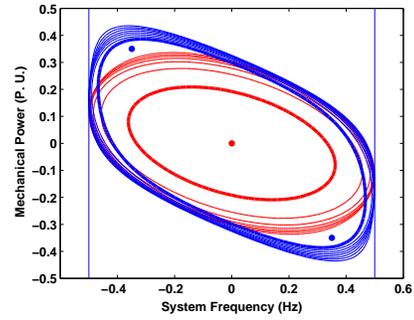}
	\caption{Iteration in calculating ROS with different initializations.}
	\label{fig_ROS_CONV}
\end{figure}
\begin{figure}[h]
	\centering
	\includegraphics[scale=0.5]{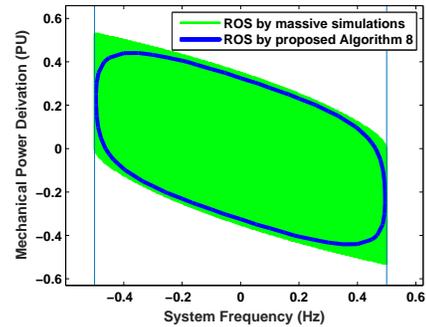}
	\caption{ROS of Mode 1 under normal condition obtained by proposed Algorithm 8 and extensive simulations.}
	\label{fig_ROS1}
\end{figure}

With the given safety limit, the ROS calculation for Mode 1 under no disturbance can be projected onto the plane $\Delta\omega$-$\Delta P_{m}$ as illustrated in Fig. \ref{fig_ROS_CONV} with two different initializations. The iteration sequence indicates that if more initial guess points are used, the fewer iterations needed and a better estimation can be achieved (as shown in the blue case). The final result is shown in Fig. \ref{fig_ROS1}. The green region is the ROS obtained by extensive simulations and can be regarded as the true one. The comparison shows that the proposed algorithm successfully reduces conservatism in the estimate for the corner effect in the polynomial-based set study.
\begin{figure*}[b]
	\centering
	\begin{minipage}[]{0.32\textwidth}
		\centering
		\includegraphics[scale=0.4]{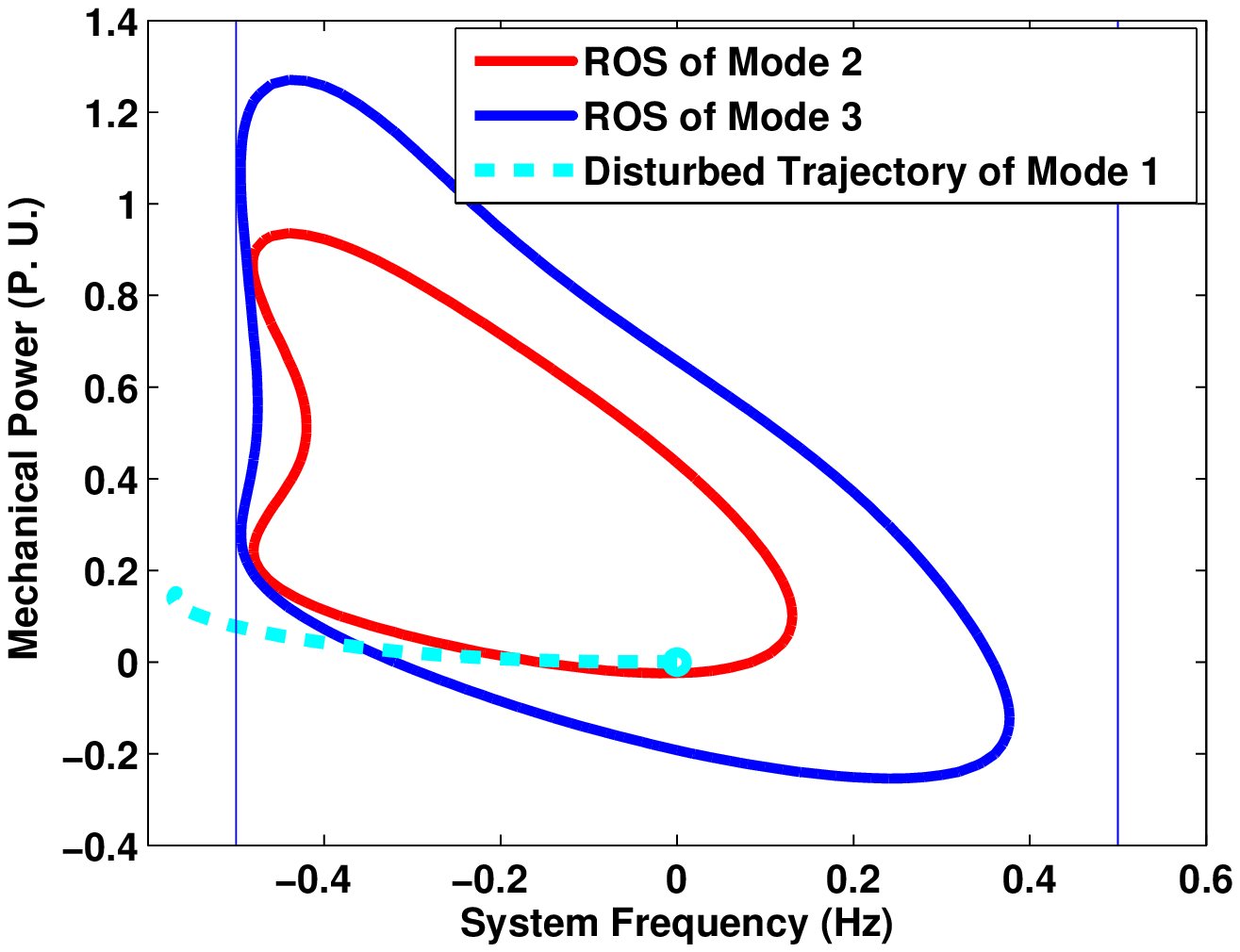}
		\caption{ROS of Mode 2 and 3 under given scenario.}\label{fig_ROS23_D15}
	\end{minipage}\hfill
	\begin{minipage}[]{0.32\textwidth}
		\centering
		\includegraphics[scale=0.4]{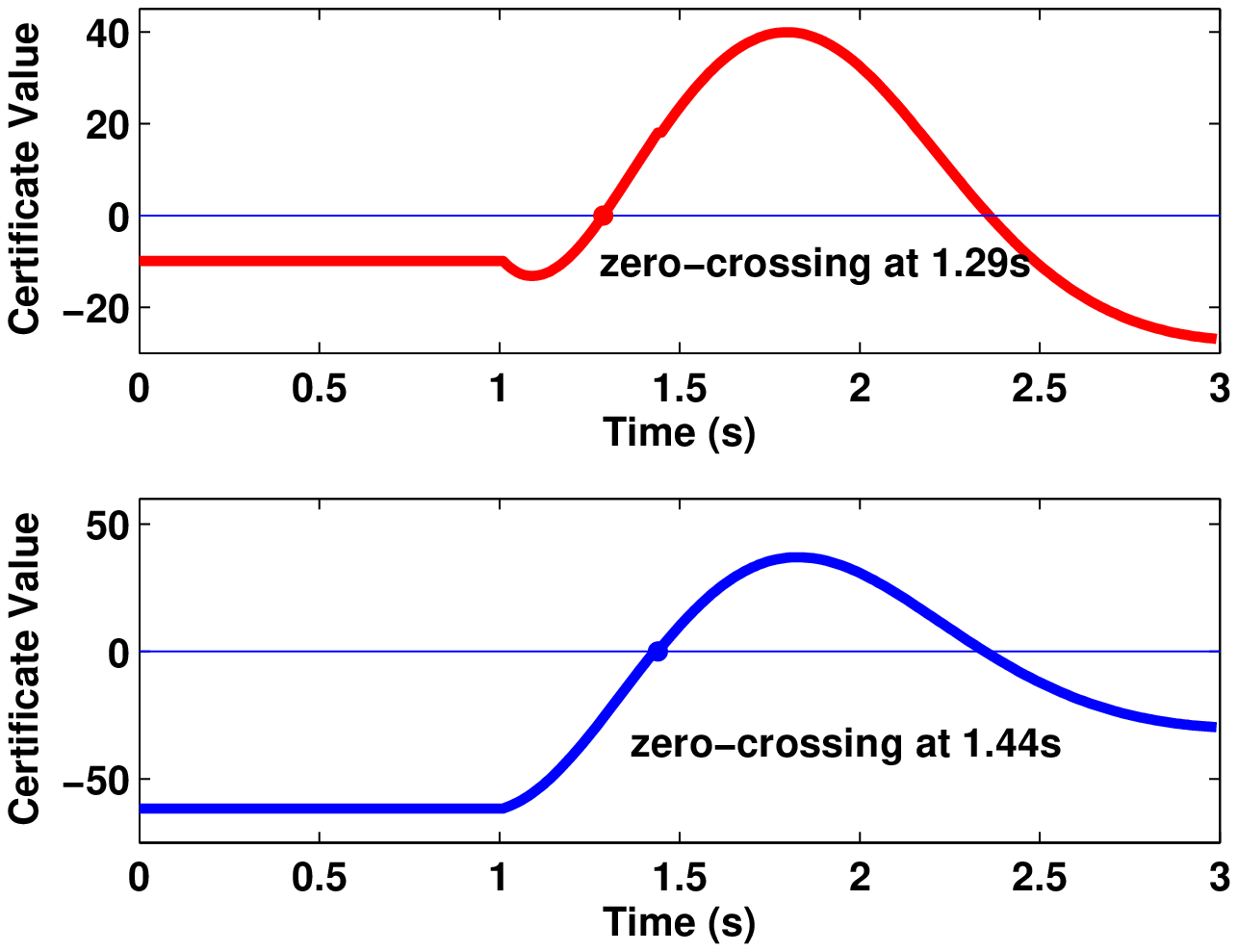}
		\caption{Value of $B_{d2}(x)$ (upper) and $B_{d3}(x)$ (lower) w.r.t  the disturbed trajectory $X_{d1}$}
		\label{fig_BV23To1}
	\end{minipage}\hfill
	\begin{minipage}[]{0.32\textwidth}
		\centering
		\includegraphics[scale=0.4]{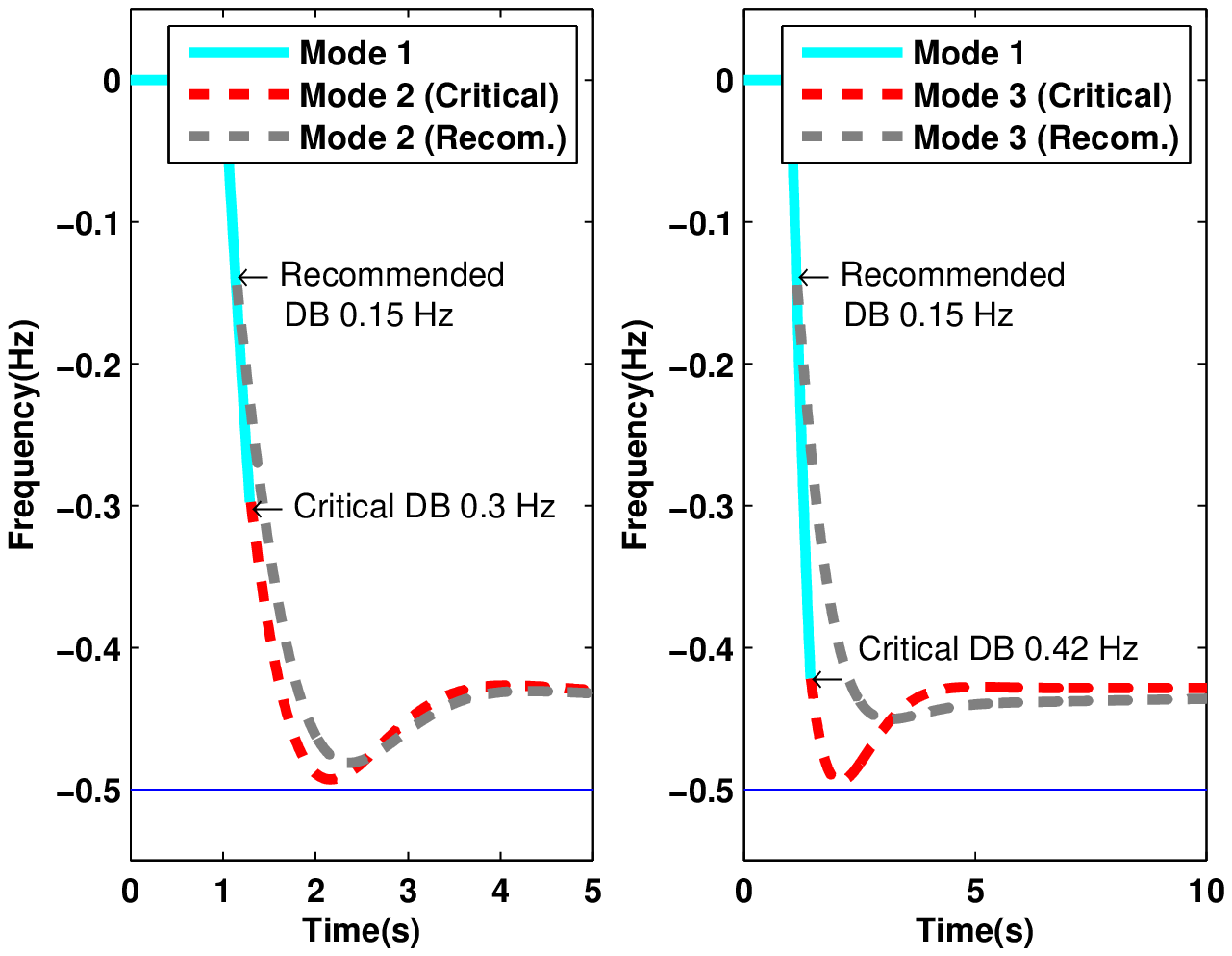}
		\caption{Frequency dynamics of full-order model under calculated critical deadband.}
		\label{fig_DB23}
	\end{minipage}
\end{figure*}
\begin{figure*}[htbp!]
	\centering
	\begin{minipage}[]{0.32\textwidth}
		\centering
		\includegraphics[scale=0.4]{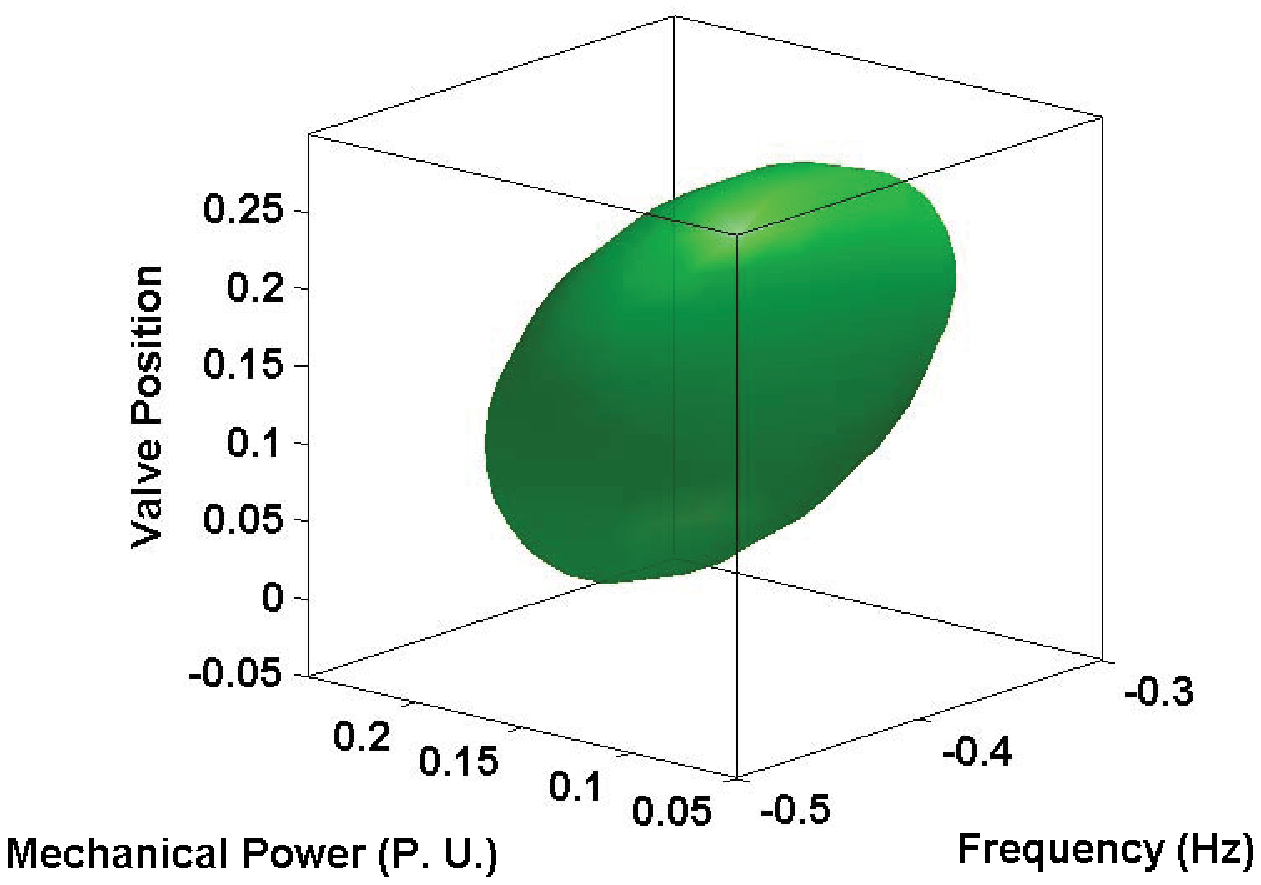}
		\caption{ROS of Mode 1 under the given scenario.}\label{fig_ROS1_D15}
	\end{minipage}\hfill
	\begin{minipage}[]{0.32\textwidth}
		\centering
		\includegraphics[scale=0.4]{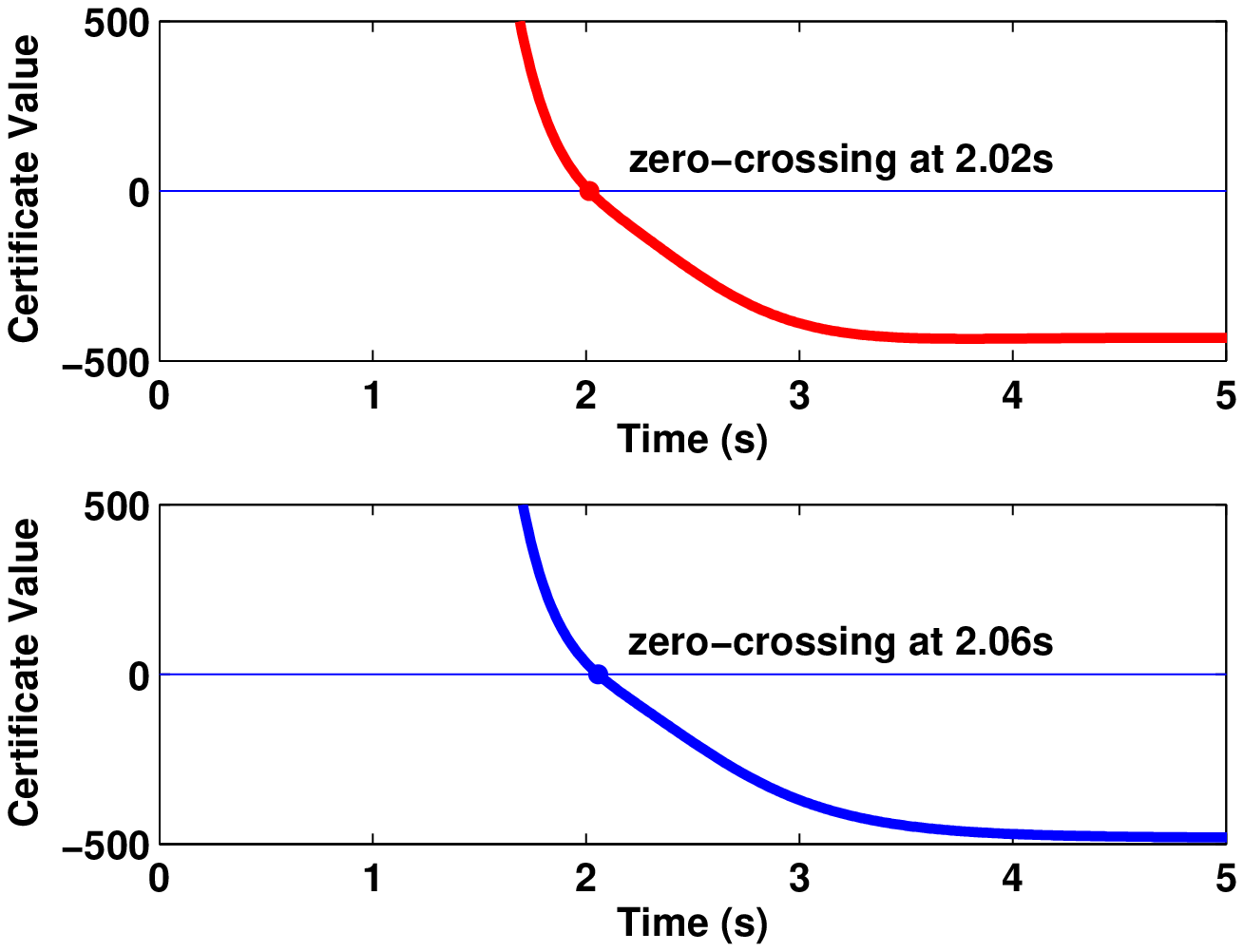}
		\caption{Value of $B_{d1}(x)$ w.r.t. the disturbed trajectories $X_{d12}$ (upper) and $X_{d13}$ (lower).}
		\label{fig_BV1WRT23}
	\end{minipage}\hfill
	\begin{minipage}[]{0.32\textwidth}
		\centering
		\includegraphics[scale=0.4]{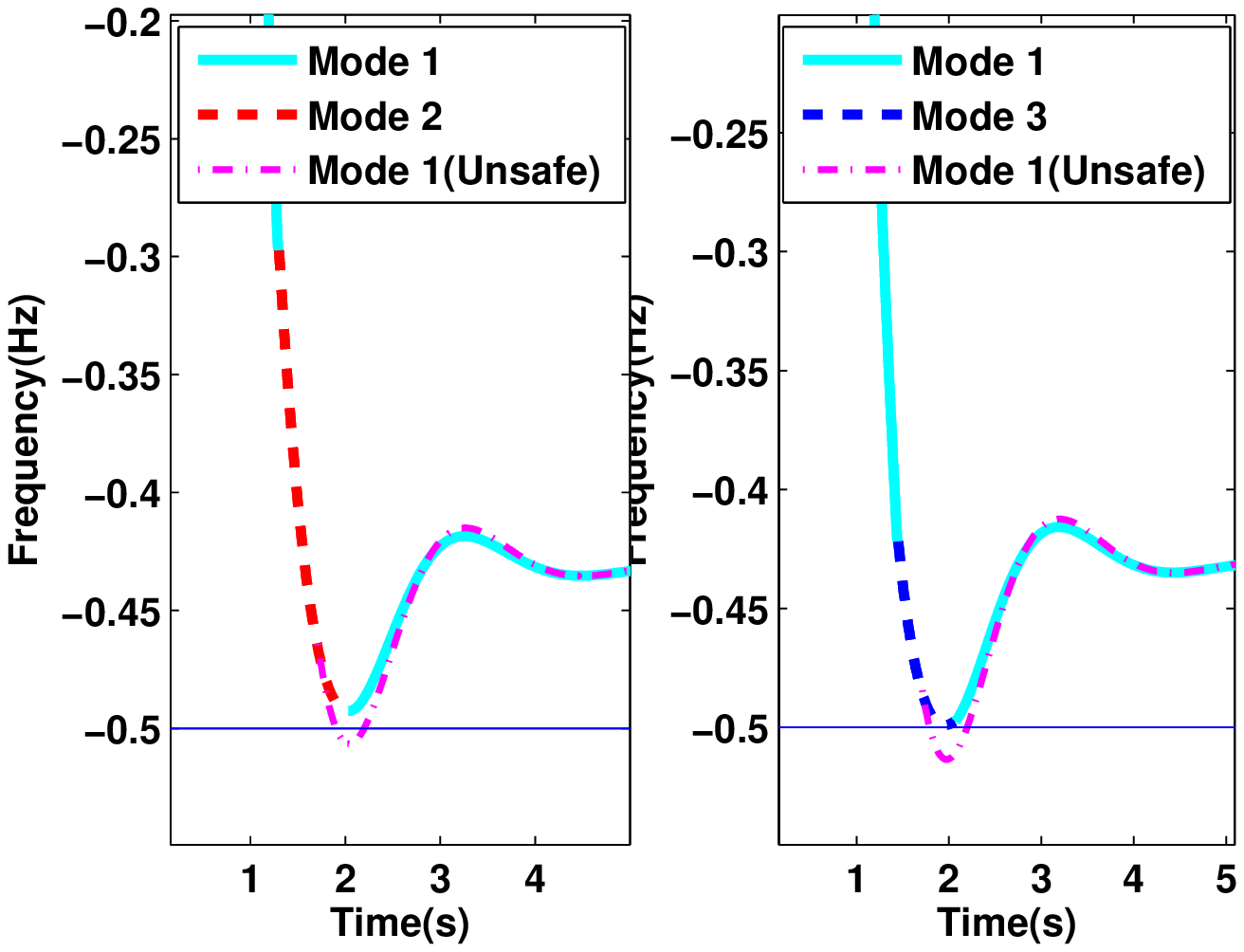}
		\caption{Frequency dynamics under critical deadband and threshold.}
		\label{fig_TH23back1}
	\end{minipage}
\end{figure*}
\subsection{IE Mode Only}
\label{subsec_IE}
The ROS under the worst-case scenario of Modes 1-3 are calculated with representation of polynomials in terms of $x_{\text{rd}}$ up to degree 8. Denote these regions as
\begin{align*}
\text{Worst-case ROS 1: }\mathcal{S}_{d1}=\left\lbrace x_{\text{rd}}:B_{d1}(x_{\text{rd}})\leq 0 \right\rbrace \\
\text{Worst-case ROS 2: }\mathcal{S}_{d2}=\left\lbrace x_{\text{rd}}:B_{d2}(x_{\text{rd}})\leq 0 \right\rbrace \\
\text{Worst-case ROS 3: }\mathcal{S}_{d3}=\left\lbrace x_{\text{rd}}:B_{d3}(x_{\text{rd}})\leq 0 \right\rbrace 
\end{align*}
where
$B_{d1}(x_{\text{rd}})$, $B_{d2}(x_{\text{rd}})$ and $B_{d3}(x_{\text{rd}})$ serve as safety switching guards.\par
To determine if the safety limits can be preserved, one needs to check whether the intersection between $\mathcal{S}_{d1}$ and the pre-disturbed operating point $x_{0}$ is empty. In our case, the fact that $\mathcal{S}_{d1}\cap \{x_{0}\}=\varnothing $ is graphically shown in Fig. \ref{fig_ROS1_D15} and mathematically verified by $B_{d1}(x_{0})>0$. According to Proposition \ref{thm_ros_hybrid_principle}, the safety of frequency cannot be preserved without inertia emulation as shown in Fig \ref{fig_FD_Modes}.\par
To verify the largest deadband or equivalently the critical switching instant from Mode 1 to Mode 2 or 3, the values of $B_{d2}(x_{\text{rd}})$ and $B_{d3}(x_{\text{rd}})$ with respect to the disturbed trajectory of Mode 1, denoted as $X_{d1}$ (dash line in Fig. \ref{fig_ROS23_D15}), is calculated. Note that $X_{d1}$ is from the full-order model and only relevant states $\bar{X}_{d1}=[X_{d1}(1),X_{d1}(2),X_{d1}(3),X_{d1}(6)]$ are substituted into the guards. The zero-crossing point from negative to positive values denotes the critical switching instant $t_{\text{cr}}$, or equivalently largest deadband with the value $\Delta\omega(t_{\text{cr}})$. As shown in Fig. \ref{fig_BV23To1}, the largest deadband (critical switching instant) is 0.30 Hz (1.29 s) if Mode 2 is used, and 0.42 Hz (1.44 s) if Mode 3 is used. Simulation of each scenario with the suggested largest deadband as well as the recommended value from GE (0.15 Hz) is carried out and shown in Fig. \ref{fig_DB23}. As seen the system safety is preserved, but the recommended values are conservative especially when Mode 3 is activated. On the other hand, the fact that the largest frequency excursion point is extremely close to the limit indicates that the estimated ROS is not overly conservative.\par
Beyond safety, the earliest support deactivation (earliest switching instant) is established so that the emulated inertia can be shed to obtain faster frequency restoration. Let the trajectories in Fig. \ref{fig_DB23} be denoted as $X_{d12}$ (left) and $X_{d13}$ (right) and substituted into $B_{d1}(x_{\text{rd}})$ to find the zero-crossing point from positive to negative, which occurs approximately at 2 s for both cases. Negativity of the safety switching guard $B_{d1}(\bar{X}_{d1i})$ guarantees safe switching from Mode $i$ to Mode 1. Early switching when $B_{d1}(\bar{X}_{d1i})>0$ will lead to an unsafe trajectory. Both cases are shown in Fig. \ref{fig_TH23back1}.
\subsection{IEPFC Mode and Safety Recovery}
\label{subsec_IEPFC}
\begin{figure}[htbp!]
	\centering
	\includegraphics[scale=0.35]{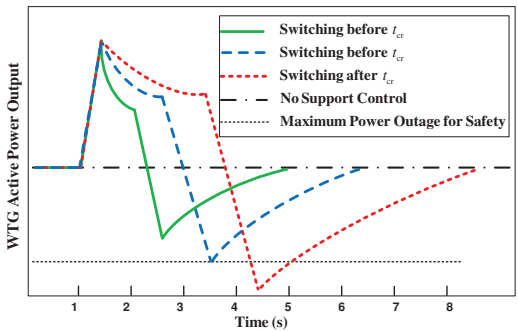}
	\caption{An equal-area criterion in support-deactivation procedure.}
	\label{fig_Windows_Principle}
\end{figure}

The additional PFC loop will artificially create additional load-frequency sensitivity and the maximum frequency excursion will decrease. The deadband analysis procedure is similar and will not be repeated. However, when it comes to support deactivation, due to the additional constant frequency deviation, a safe switching time window appears. Thus, the PFC mode needs to be deactivated before a critical time. The mechanism is illustrated in Fig. \ref{fig_Windows_Principle}.\par
\begin{figure}[htbp!]
	\centering
	\includegraphics[scale=0.4]{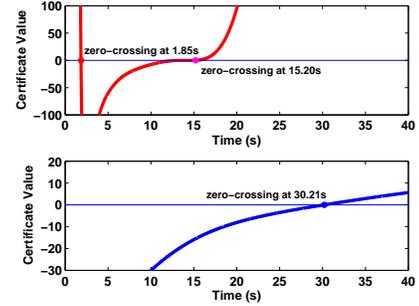}
	\caption{Upper: Value of $B_{d1}(x)$ and $B_{d3}(x)$ w.r.t. the disturbed trajectory $X_{d15}$.}
	\label{fig_window}
\end{figure}
\begin{figure}[htbp!]
	\centering
	\includegraphics[scale=0.4]{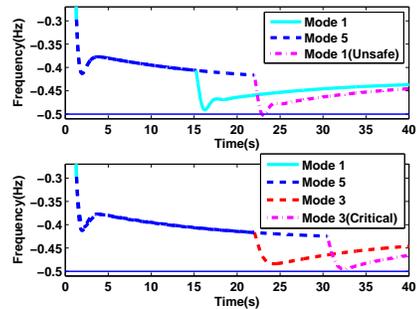}
	\caption{Frequency dynamics under 0.3 Hz deadband and critical deactivation: normal sequence Mode 1-5-1 (Upper) and safe recovery sequence Mode 1-5-3-1 (lower)}
	\label{fig_SafeRecover}
\end{figure}

The WTG attempts to draw the total energy that is pulled out during support mode to restore the rotor speed so the upper area and lower areas have to be equal. When the PFC is deactivated, $P_{\text{gen}}$ decreases to satisfy this equal area criterion. This sudden shortage of active power, if large enough, will lead to an unsafe trajectory. Let Mode 5 be designed with a deadband of 0.3 Hz and substitute the disturbed trajectory $\bar{X}_{d15}$ into $B_{d1}(x_{\text{rd}})$, then the critical switching window can be observed as in the upper plot of Fig. \ref{fig_window}, where the critical deactivating instants suggested by the guard is 15.2 s. A deactivation at 22 s leads to an unsafe trajectory. Frequency dynamics of both cases are shown in the upper plot of Fig. \ref{fig_SafeRecover}.\par
To extend the critical deactivating instant, we propose a safety recovery procedure illustrated by Fig. \ref{fig_deadband_hybrid_model}. When the PFC mode is deactivated, the corresponding IE mode is kept to manage the sudden shortage of active power. By checking the value of $B_{d3}(x_{\text{rd}})$ with respect to the trajectory $\bar{X}_{d15}$ (lower plot of Fig. \ref{fig_window}) this procedure extends the critical deactivating instant by 15 s. The original unsafe switching from Mode 5 directly to Mode 1 at 22 s is now safely switched to Mode 3 as shown in the lower plot of Fig. \ref{fig_SafeRecover}. The critical switching instant from Mode 5 to 3 is suggested to be 30.21 s by the guard in Fig. \ref{fig_window} and verified by simulation in Fig. \ref{fig_SafeRecover}.\par
\par
\section{Conclusion}
\label{sec_con}
This paper proposes a principle of WTG mode switching synthesis for safe frequency response. Given desired safety limits of grid frequency and worst-case scenarios, an SOS optimization based algorithm is developed to enlarge the estimation of ROS, and then the critical switching instants, equivalent to largest deadband, for safety preservation are obtained. Simulation shows that the critical switching instants are not overly conservative. A switching sequence for safe recovery of WTG rotor speed is proposed as well. In addition, the emulated inertia and load-damping effect is derived in the time frame of inertia and primary frequency response, respectively.\par
Future study will focus on analyzing larger power networks. From a methodology point of view, instead of choosing monomials as a dense basis to represent the set of continuous functions on compact sets, other representations such as the Handelman representation can be employed, which will lead to linear relaxations of polynomial positivity rather than SOS relaxations \cite{sassi2015linear}. Coordination of multiple renewable sources as a redundant actuator set under a severe contingency can be considered as well \cite{Ehsan}.

\appendices
\section{System Parameters and Operating Condition}
\label{appendix_data}
\emph{Type-3 wind turbine generator parameters}\par
\noindent$X_{m}=3.5092$, $X_{s}=3.5547$, $X_{r}=3.5859$, $R_{s}=0.01015$, $R_{r}=0.0088$, $H_{D}=4(\text{s})$, $p=4$, $\rho=1.225(\text{kg/m}^{3})$, $R_{t}=38.5(\text{m})$, $S_{bD}=1 (\text{MVA})$, $C_{\text{opt}}=3.2397\times 10^{-7}(\text{s}^{3}/\text{Hz}^{3})$, $k=1/45$, $K_{P1}=K_{P2}=K_{P3}=K_{P4}=1$, $K_{I1}=K_{I2}=K_{I3}=K_{I4}=5$.\par
\emph{System frequency response model parameters}\par
\noindent$\omega_{s}=60(\text{Hz})$, $D=1$, $H=4(\text{s})$, $\tau_{\text{ch}}=0.3(\text{s})$, $\tau_{g}=0.1(\text{s})$, $R=0.05$.\par
\emph{Network parameters and operating condition}\par
\noindent$X_{t}=0.07$, $\bar{V}=1$, $\bar{\theta}=0(\text{rad})$, $\bar{\theta}_{t}=0(\text{rad})$, $\bar{v}_{\text{wind}}=12(\text{m/s})$, $\bar{\omega}_{r}=72(\text{Hz})$, $\bar{P}_{\text{gen}}=0.3$, $\bar{Q}_{\text{set}}=0$, $S_{b}=1000(\text{MVA})$
\section{Wind Turbine Generator Model}
\label{appendix_all_model}
The standard wind turbine model is given as follows \cite{hector}
\begin{subequations}
\label{eq_turbine}
\begin{align}
\lambda=& \frac{2k\omega_{r}R_{t}}{pv_{\text{wind}}}\\
\lambda_{i}=& \left(\frac{1}{\lambda+0.08\theta_{t}}-\frac{0.035}{\theta_{t}^{3}+1} \right)^{-1}\\
C_{p}=& 0.22\left(\frac{116}{\lambda_{i}}-0.4\theta_{t}-5\right) e^{-\frac{12.5}{\lambda_{i}}}\\
T_{m}=&\frac{1}{2}\frac{\rho\pi R_{t}^{2}\omega_{b}C_{p}v_{\text{wind}}^{3}}{S_{b}\omega_{r}}
\end{align}
\end{subequations}\par
The type-3 wind turbine generator differential equations are given as follows \cite{hector}
\begin{subequations}
\label{eq_WTG_diff}
\begin{align}
\begin{split}
\dot{E}^{\prime}_{qD}=&-\frac{1}{T^{\prime}_{0}}(E^{\prime}_{qD}+(X_{s}-X^{\prime}_{s})I_{ds})+\omega_{s}\dfrac{X_{m}}{X_{r}}V_{dr} \\
&-(\omega_{s}-\omega_{r})E^{\prime}_{dD} \end{split}\\
\begin{split}
\dot{E}^{\prime}_{dD}=&-\frac{1}{T^{\prime}_{0}}(E^{\prime}_{dD}+(X_{s}-X^{\prime}_{s})I_{qs})+\omega_{s}\dfrac{X_{m}}{X_{r}}V_{qr}\\
&-(\omega_{s}-\omega_{r})E^{\prime}_{qD}  \end{split}\\
\dot{\omega}_{r}=&\frac{\omega_{s}}{2H_{D}}(T_{m}-E^{\prime}_{dD}I_{ds}-E^{\prime}_{qD}I_{qs})\\
\dot{x}_{1}=& K_{I1}(P_{\text{ref}}-P_{\text{gen}})\\
\dot{x}_{2}=& K_{I2}(K_{P1}(P_{\text{ref}}-P_{\text{gen}})+x_{1}-I_{qr})\\
\dot{x}_{3}=& K_{I3}(Q_{\text{ref}}-Q_{\text{gen}})\\
\dot{x}_{4}=& K_{I4}(K_{P3}(Q_{\text{ref}}-Q_{\text{gen}})+x_{3}-I_{dr})
\end{align}
\end{subequations}
where $E^{\prime}_{dD}$, $E^{\prime}_{qD}$ and $\omega_{r}$ are $d$ $q$ axis voltage and rotor speed of wind turbine generator, respectively. $x_{1}$ to $x_{4}$ are proportional-integral regulator induced states. And $P_{\text{ref}}=C_{\text{opt}}\omega_{r}^{3}$, $Q_{\text{ref}}=Q_{\text{set}}$, $T^{\prime}_{0}=\frac{X_{r}}{\omega_{s}R_{r}}$ and $X_{s}^{\prime}=X_{s}-\frac{X_{m}^{2}}{X_{r}}$.\par
The type-3 wind turbine generator algebraic equations are given as follows \cite{hector}
\begin{subequations}
\label{eq_WTG_alg}
\begin{align}
\begin{split}
0 =& K_{P2}(K_{P1}(P_{\text{ref}}-P_{\text{gen}})+x_{1}-I_{qr})\\&+x_{2}-V_{qr} \end{split}\\
\begin{split}
0 =& K_{P4}(K_{P3}(Q_{\text{ref}}-Q_{\text{gen}})+x_{3}-I_{dr})\\&+x_{4}-V_{dr} \end{split}\\
\begin{split}
0 =&-P_{\text{gen}}+E^{\prime}_{dD}I_{ds}+E^{\prime}_{qD}I_{qs}-R_{s}(I^{2}_{ds}+I^{2}_{qs})\\&-(V_{qr}I_{qr}+V_{dr}I_{dr}) \end{split}\\
0 =&-Q_{\text{gen}}+E^{\prime}_{qD}I_{ds}+E^{\prime}_{dD}I_{qs}-X^{\prime}_{s}(I^{2}_{ds}+I^{2}_{qs})\\
0 =&-I_{dr}+\frac{E^{\prime}_{qD}}{X_{m}}+\frac{X_{m}}{X_{r}}I_{ds}\\
0 =&-I_{qr}-\frac{E^{\prime}_{dD}}{X_{m}}+\frac{X_{m}}{X_{r}}I_{qs}
\end{align}
\end{subequations}
where $V_{dr}$, $V_{qr}$, $I_{dr}$, $I_{qr}$ are rotor $d$ $q$ axis voltage and current, respectively. $V_{ds}$, $V_{qs}$, $I_{ds}$, $I_{qs}$ are stator $d$ $q$ axis voltage and current. $P_{\text{gen}}$ and $Q_{\text{gen}}$ are WTG active and reactive power output. $V_{D}$ and $\theta_{D}$ are voltage magnitude and angle of the bus which WTGs are connected to.\par
The network algebraic equations are given as follows
\begin{subequations}
\label{eq_network_alg}
\begin{align}
E^{\prime}_{qD}-jE^{\prime}_{dD}=(R_{s}+jX_{s}^{\prime})(I_{qs}-jI_{ds})+V_{D}\\
V_{D}e^{j\theta_{D}}=jX_{t}(I_{qs}-jI_{ds}-I_{GC})e^{j\theta_{D}}+Ve^{j\theta}
\end{align}
\end{subequations}
where
\begin{align*}
I_{GC}=\dfrac{V_{dr}I_{qr}+V_{dr}I_{dr}}{V_{D}}
\end{align*}

\section*{Acknowledgment}
This work was supported in part by the National Science Foundation under grant No CNS-1239366, in part by the National Science Foundation under grant No ECCS-1509114, and in part by the Engineering Research Center Program of the National Science Foundation and the Department of Energy under NSF Award Number EEC-1041877.

\bibliography{IEEEabrv_zyc,Ref}  % name your BibTeX data base
\bibliographystyle{IEEEtran}

\end{document}